\documentclass[fleqn,usenatbib]{mnras}

\usepackage[T1]{fontenc}

\DeclareRobustCommand{\VAN}[3]{#2}
\let\VANthebibliography\thebibliography
\def\thebibliography{\DeclareRobustCommand{\VAN}[3]{##3}\VANthebibliography}

\newcommand\lal{Ly-$\alpha$ }
\newcommand\lab{Ly-$\beta$ }
\newcommand\laa{Ly-$\alpha$}
\newcommand\lb{Ly-$\beta$}

\defcitealias{Bosman21}{B21}

\usepackage[table]{xcolor}
\newcommand\revi[1]{{#1}} 
\newcommand\revmath[1]{{#1}}

\usepackage{graphicx}	
\usepackage{amsmath}	
\usepackage{amssymb}	

\title[Ly-$\alpha$ optical depth with XQR-30]{Hydrogen reionisation ends by $z=5.3$: Lyman-$\alpha$ optical depth measured by the XQR-30 sample}

\author[S. E. I. Bosman et al.]{
\href{https://orcid.org/0000-0001-8582-7012}{Sarah E.~I.~Bosman}$^{1}$\thanks{E-mail: bosman@mpia.de},
\href{https://orcid.org/0000-0003-0821-3644}{Frederick B.~Davies}$^{1}$,
\href{https://orcid.org/0000-0003-2344-263X}{George D.~Becker}$^{2}$,
\href{https://orcid.org/0000-0001-5211-1958}{Laura C.~Keating}$^{3}$, \newauthor
\href{https://orcid.org/0000-0002-3324-4824}{Rebecca L.~Davies}$^{4,5}$,
\href{https://orcid.org/0000-0003-3307-7525}{Yongda Zhu}$^{2}$,
\href{https://orcid.org/0000-0003-2895-6218}{Anna-Christina Eilers}$^{6}$\textdagger,
\href{https://orcid.org/0000-0003-3693-3091}{Valentina D'Odorico}$^{7,8}$,\newauthor
\href{https://orcid.org/0000-0002-1620-0897}{Fuyan Bian}$^{9}$,
\href{https://orcid.org/0000-0002-4314-021X}{Manuela Bischetti}$^{7,10}$,
\href{https://orcid.org/0000-0002-2115-5234}{Stefano V.~Cristiani}$^{7}$,
\href{https://orcid.org/0000-0003-3310-0131}{Xiaohui Fan}$^{11}$,\newauthor
\href{https://orcid.org/0000-0002-6822-2254}{Emanuele P.~Farina}$^{12}$,
\href{https://orcid.org/0000-0001-8443-2393}{Martin G.~Haehnelt}$^{13,14}$,
\href{https://orcid.org/0000-0002-7054-4332}{Joseph F.~Hennawi}$^{15,16}$
\href{https://orcid.org/0000-0001-5829-4716}{Girish Kulkarni}$^{17}$,\newauthor
\href{https://orcid.org/0000-0003-3374-1772}{Andrei Mesinger}$^{8}$,
\href{https://orcid.org/0000-0001-5492-4522}{Romain A.~Meyer}$^{1}$,
\href{https://orcid.org/0000-0003-2984-6803}{Masafusa Onoue}$^{1}$,
\href{https://orcid.org/0000-0002-7129-5761}{Andrea Pallottini}$^{7}$,\newauthor 
\href{https://orcid.org/0000-0002-4314-1810}{Yuxiang Qin}$^{18,5}$,
\href{https://orcid.org/0000-0002-5360-8103}{Emma Ryan-Weber}$^{4,5}$,
\href{https://orcid.org/0000-0002-4544-8242}{Jan-Torge Schindler}$^{1,16}$,
\href{https://orcid.org/0000-0003-4793-7880}{Fabian Walter}$^{1}$,\newauthor
\href{https://orcid.org/0000-0002-7633-431X}{Feige Wang}$^{11}$\textdagger, 
\href{https://orcid.org/0000-0001-5287-4242}{Jinyi Yang}$^{11}$
\\
$^{1}$ Max-Planck-Institut f{\"u}r Astronomie, K{\"o}nigstuhl 17, D-69117 Heidelberg, Germany \\
$^{2}$ Department of Physics \& Astronomy, University of California, Riverside, CA, 92521, USA \\
$^{3}$ Leibniz-Institut fur Astrophysik Potsdam, An der Sternwarte 16, Potsdam 14482, Germany\\
$^{4}$ Centre for Astrophysics and Supercomputing, Swinburne University of Technology, Hawthorn, Victoria 3122, Australia \\
$^{5}$ ARC Centre of Excellence for All Sky Astrophysics in 3 Dimensions (ASTRO 3D), Australia \\
$^{6}$ MIT Kavli Institute for Astrophysics and Space Research, 77 Massachusetts Ave., Cambridge, MA02139, USA\\
$^{7}$ INAF-Osservatorio Astronomico di Trieste, Via Tiepolo 11, I-34143 Trieste, Italy \\
$^{8}$ Scuola Normale Superiore, Piazza dei Cavalieri 7, 56126 Pisa, Italy \\
$^{9}$ European Southern Observatory, Alonso de C\'ordova 3107, Casilla 19001, Vitacura, Santiago 19, Chile \\
$^{10}$ INAF  -  Osservatorio  Astronomico  di  Roma,  Via  Frascati  33,  I--00078 Monte Porzio Catone, Italy\\
$^{11}$ Steward Observatory, University of Arizona, 933 North Cherry Avenue, Tucson, AZ 85721, USA\\
$^{12}$ Max Planck Institut fur Astrophysik, Karl--Schwarzschild--Stra\ss e 1, D-85748 Garching bei M\"unchen, Germany\\
$^{13}$ Institute of Astronomy, University of Cambridge, Madingley Road, Cambridge CB3 0HA, UK\\
$^{14}$ Kavli Institute for Cosmology, University of Cambridge, Madingley Road, Cambridge CB3 0HA, UK\\
$^{15}$ Department of Physics, Broida Hall, University of California, Santa Barbara Santa Barbara, CA 93106-9530, USA\\
$^{16}$ Leiden Observatory, Leiden University, PO Box 9513, NL-2300 RA Leiden, the Netherlands\\
$^{17}$ Department of Theoretical Physics, Tata Institute of Fundamental Research, Homi Bhabha Road, Mumbai 400005, India\\
$^{18}$ School of Physics, University of Melbourne, Parkville, VIC 3010, Australia\\
\textdagger NASA Hubble Fellow\\
}

\date{}

\pubyear{2021}

\usepackage{newtxtext,newtxmath}

\begin{document}
\label{firstpage}
\pagerange{\pageref{firstpage}--\pageref{lastpage}}
\maketitle

\begin{abstract}

The presence of excess scatter in the \lal forest at $z\sim 5.5$, together with the existence of sporadic extended opaque Gunn-Peterson troughs, has started to provide robust evidence for a late end of hydrogen reionisation. However, low data quality and systematic uncertainties complicate the use of \lal transmission as a precision probe of reionisation's end stages. 
In this paper, we assemble a sample of $67$ quasar sightlines at $z>5.5$ with high signal-to-noise ratios of $>10$ per $\leq 15$ km s$^{-1}$ spectral pixel, relying largely on the new XQR-30 quasar sample. XQR-30 is a large program on VLT/X-Shooter which obtained deep (SNR$>20$ per pixel) spectra of $30$ quasars at $z>5.7$. We carefully account for systematics in continuum reconstruction, instrumentation, and contamination by damped \lal systems. We present improved measurements of the mean \lal transmission over $4.9<z<6.1$. Using all known systematics in a forward modelling analysis, we find excellent agreement between the observed \lal transmission distributions and the homogeneous-UVB simulations Sherwood and Nyx up to $z\leq5.2$ ($<1\sigma$), and mild tension ($\sim2.5\sigma$) at $z=5.3$. Homogeneous UVB models are ruled out by excess \lal transmission scatter at $z\geq5.4$ with high confidence ($>3.5\sigma$). Our results indicate that reionisation-related fluctuations, whether in the UVB, residual neutral hydrogen fraction, and/or IGM temperature, persist in the intergalactic medium \revi{until at least} $z=5.3$ ($t=1.1$ Gyr after the Big Bang). This is further evidence for a late end to reionisation. 
\end{abstract}




\section{Introduction}

The epoch of reionisation, during which the bulk of intergalactic hydrogen became ionised, is of great interest for both astrophysicists and cosmologists. The timing and morphology of the transition relate to the properties of the first galaxies and other potential reionising sources, holding crucial information on the large-scale properties of the intergalactic medium (IGM) as well as galaxy formation and evolution at early cosmic times (see e.g.~\citealt{Dayal18}). 
Next-generation 21cm experiments aim to directly detect the signature of neutral gas in the first stages of reionisation at $z\gtrsim10$ within the coming decade \citep{HERA,Trott19}.  Meanwhile, the end stages of reionisation at $z\lesssim7$ are already being probed through quasar absorption in the Lyman-$\alpha$ (\laa) and Lyman-$\beta$ (\lb) hydrogen transitions (e.g.~\citealt{Fan02,Mesinger04,Mortlock11,Bosman15,Greig17,Davies18-DW,Eilers19, Wang20}). 

Over $\revmath{400}$ quasars are now known at $z>5.7$, corresponding to the first billion years after the Big Bang \citep{list}. The first observational constraints on the end of reionisation originated from detections of Gunn-Peterson (GP; \citealt{GP}) troughs at $z>6$: total absorption of quasar continuum emission by neutral hydrogen in the IGM \citep{Fan00, Fan06}. Saturation of \lal absorption occurs in the presence of IGM gas with a hydrogen neutral fraction $\gtrsim0.01\%$, with dependence on the density and temperature of the gas. The interpretation of GP troughs for reionisation is complex. 
Measurements of \lal transmission towards quasars have revealed that saturation occurs sporadically down to $z\sim5.6$, and also on very large contiguous scales $\gtrsim 100$ cMpc/h \citep{Becker15}. Observed differences in \lal optical depth between sightlines at fixed redshift far exceed expectations from cosmic density fluctuations alone \citep{Becker15, Bosman18, Eilers18, Yang20}, implying a more protracted or `patchy' end of reionisation than was unforeseen by standard models (but see \citealt{Lidz06, Mesinger10}).

Determining the nature of these $z<6$ optical depth fluctuations is currently a major goal of reionisation theory. The existence of late-persisting GP troughs and the observed optical depth scatter at $z\sim 5.8$ can be matched by a late end of reionisation in which some voids with hydrogen neutral fractions $>10\%$ persist down to $z\sim 5.6$ \citep{Kulkarni19, Keating20, Nasir20}. Roughly half of the cosmic volume would then be occupied by neutral gas at $z\sim 7$, with important consequences such as e.g.~facilitating the observation of the 21cm signal \citep{Raste21, Soltinsky21}. 
In addition, scatter in the \lal optical depth at $z \sim 5.8$ can also arise from a short and fluctuating photon mean free path, which alters the propagation of ionising photons through the IGM \citep{Davies16, Daloisio18}. Recent observations have suggested a shorter-than-expected ionising mean free path at $z\sim 6$ \revi{(\citealt{Becker21}; see also \citealt{Bosman21-mfp})}. While not \revi{explicitly} requiring a late end to reionisation, a short mean free path at $z\sim6$ poses tight requirements on the ionising power of early galaxies \citep{Davies21, Cain21}. 
Further models have explored the importance of additional sources of scatter, such as relic IGM temperature fluctuations \citep{Daloisio15, Keating18} or a potential significant role of quasars \citep{Chardin17, Meiksin20}. 
Meanwhile, observations of \lal transmission at $z\lesssim 5$ are fully consistent with IGM models including only the effects of density fluctuations in a homogeneous (i.e.~permeated) ultra-violet background (UVB) \citep{Rollinde13, Becker13, Becker15}. The transition between these two regimes across $5.0\leq z \leq 5.8$ therefore holds crucial clues to the changes in IGM properties as reionisation finishes.

Furthermore, the first measurements of the \lal optical depth distribution at $z\sim5.8$ with large quasar samples have lead to the first results from semi-numerical models of reionisation's patchy end stages. The Bayesian inference enabled by these semi-numerical models allowed us to statistically constrain the end of reionisation to $z<5.6$ \citep{Choudhury21, Qin21}, as well as disfavor a strong evolution of the ionising escape fraction in reionising galaxies \citep{Qin21}.





The advent of expensive and specifically tuned simulations, as well as sensitive inference models of reionisation, necessitate that measurements of \lal optical depth have a firm grasp on possible observational biases. To capture cosmic variance, studies require very large samples of quasars -- but until now, this has come at the expense of data homogeneity and potential instrumental and reduction biases which are not known accurately \citep{Bosman18}. There has been some tension between results from different groups \citep{Bosman18, Eilers18, Yang20} which can be  largely attributed to differing choices of methods for reconstructing the underlying quasar emission (\citealt{Bosman21}, thereafter \citetalias{Bosman21}). Out of necessity (insufficient data) and low relative importance compared to sample size, a rigorous quantitative examination of those biases and uncertainties has been neglected until now. Indeed, the existence of excess scatter in \lal optical depth  at $z\sim5.7$ is established very robustly even with the most pessimistic assumptions on measurement errors \citep{Becker15}. The rigour and precision required for quantitative inference and comparison to new models, however, requires a higher level of attention to observational biases and uncertainties. The XQR-30 sample (D'Odorico et al.~in prep) consisting of $30$ new high-SNR spectra of $z\gtrsim5.8$ quasars, enables such a careful analysis for the first time without sacrificing sample size.

In this paper, we use the XQR-30 sample together with archival spectra of equal quality to significantly refine measurements of \lal optical depth at $5.0\leq z\leq 6.0$. The observational data is described in Section \ref{sec:observations}. Restricting the analysis to high-quality data enables the suitable treatment of a slew of systematics and rigorous error estimation, which we describe in Section \ref{sec:methods}. We present the new distributions in Section \ref{sec:results}. Finally, we compare our measurements with expectations from a homogeneously-ionised Universe in Section \ref{sec:models}. The comparison to models both tests whether our analysis has accounted for all significant systematics at $z\sim5$, where no reionisation-related fluctuations are expected in \lal transmission, and quantifies the point of transition beyond which these fluctuations are detected. We summarise our results in Section \ref{sec:ccl}.

Throughout the paper we assume a \citet{Planck20} cosmology with $H_0 = 67.74, \Omega_m = 0.3089$. Wavelengths always refer to the rest-frame unless explicitly stated. Comoving and proper distances are always labelled explicitly (e.g.~cMpc).







\begin{table}
\begin{tabular}{l c c c}
Quasar ID & $z_{\text{qso}}$ & SNR pix$^{-1}$ & Refs. \\
\hline
\hline
PSO J323+12 & $ 6.5872 $ & $ 35.9 $ & (1,27)  \\
PSO J231-20 & $ 6.5869 $ & $42.3$ & (1,27) \\
VDES J0224-4711 & $ 6.5223 $ & $ 24.4 $ & (3,39) \\
PSO J1212+0505 & $ 6.4386 $ & $ 55.8 $ & (1,4) \\
DELS J1535+1943 & $ 6.3932 $ & $ 22.6 $ & (5,--) \\
ATLAS J2211-3206 & $ 6.3394 $ & $ 37.5 $ & (6/7,4) \\
PSO J060+24 & $ 6.192 $ & $ 49.7 $ & (8,--) \\
PSO J065-26 & $ 6.1871 $ & $ 77.9 $ & (8,27) \\
PSO J359-06 & $ 6.1722 $ & $ 68.8 $ & (9,40) \\
PSO J217-07 & $ 6.1663 $ & $ 33.3 $ & (8,8) \\
PSO J217-16 & $ 6.1498 $ & $ 73.0 $ & (8,4) \\
PSO J239-07 & $ 6.1102 $ & $ 56.3 $ & (8,40) \\
SDSS J0842+1218 & $ 6.0754 $ & $ 83.2 $ & (11/12,27) \\
ATLAS J158-14 & $  6.0685 $ & $ 60.3 $ & (6,40) \\
VDES J0408-5632 & $ 6.0345 $ & $ 86.6 $ & (3,3) \\
ATLAS J029-36 & $ 6.021 $ & $ 57.1 $ & (14,13) \\
SDSS J2310+1855 & $ 6.0031 $ & $ 113.4 $ & (15,16) \\
PSO J007+04 & $ 6.0015 $ & $ 54.4 $ & (12/17,27) \\
PSO J029-29 & $ 5.984 $ & $ 65.6 $ & (8,8) \\
PSO J108+08 & $ 5.9485 $ & $ 104.8 $ & (8,8) \\
PSO J183-12 & $ 5.917 $ & $ 61.8 $ & (17,--) \\
PSO J025-11 & $ 5.844 $ & $ 50.6 $ & (8,--) \\
PSO J242-12 & $ 5.837 $ & $ 22.9 $ & (8,--) \\
PSO J065+01 & $ 5.833 $ & $ 25.1 $ & (2,--) \\
PSO J308-27 & $ 5.7985 $ & $ 53.2 $ & (8,2) \\
\hline
\end{tabular}
\caption{XQR-30 quasars with X-Shooter spectra included in this work. References correspond to (Discovery, Redshift determination). The full list of references is given in the caption of Table~\ref{table:ESI}. } \label{table:XQR30}
\end{table}

\begin{table}
\begin{tabular}{l c c c}
Quasar ID & $z_{\text{qso}}$ & SNR pix$^{-1}$ & Refs. \\
\hline
\hline
PSO J036+03 & $ 6.5405 $ & $ 61.4 $ & (18,27)\\
PSO J011+09 & $ 6.4695 $ & $ 14.5 $ & (1,40) \\
PSO J159-02 & $ 6.386 $ & $ 22.9 $ & (8,--) \\
SDSS J0100+2802 & $ 6.3269 $ & $ 560.5 $ & (20,27) \\
ATLAS J025-33 & $ 6.318 $ & $ 127.3 $ & (14,13) \\
SDSS J1030+0524 & $ 6.309 $ & $ 69.6 $ & (21,22) \\
VDES J0330-4025 & $ 6.239 $ & $ 17.0 $ & (3,10) \\
PSO J308-21 & $ 6.2355 $ & $ 24.4 $  & (8,27)  \\
VIK J2318-3029 & $ 6.1456 $ & $ 16.5 $  & (7,27)  \\
ULAS J1319+0950 & $ 6.1347 $ & $ 81.7 $ & (23,27) \\
CFHQS J1509-1749 & $ 6.1225 $ & $ 43.0 $ & (24,4) \\
CFHQS J2100-1715 & $ 6.0807 $ & $ 12.4 $ & (25,27) \\
ULAS J1207+0630 & $ 6.0366 $ & $ 29.2 $ & (12,4)  \\
SDSS J1306+0356 & $ 6.033 $ & $ 65.3 $ &  (21,27) \\
PSO J340-18 & $ 5.999 $ & $ 29.9 $ &  (17,13) \\
ULAS J0148+0600 & $ 5.998 $ & $ 152.0 $ & (12,13) \\
SDSS J0818+1722 & $ 5.997 $ & $ 132.1 $ & (19,13) \\
VIK J0046-2837 & $ 5.9926 $ & $ 15.0 $ &  (28,29) \\
PSO J056-16 & $ 5.9676 $ & $ 32.0 $ & (8,40) \\
PSO J004+17 & $ 5.8166 $ & $ 15.9 $ & (8,40)\\
SDSS J0836+0054 & $ 5.804 $ & $ 73.8 $ & (21,--) \\
SDSS J0927+2001 & $ 5.7722 $ & $ 53.8 $ & (19,26) \\
PSO J215-16 & $ 5.7321 $ & $ 30.2 $ &  (31,31) \\
J1335-0328 & $ 5.693 $ & $ 35.0 $ & (32,13) \\
J0108+0711 & $ 5.577 $ & $ 20.0 $ & (32,13)  \\
J2207-0416 & $ 5.529 $ & $ 16.9 $ & (9,13) \\

\hline
\end{tabular}
\caption{Quasars with literature and archival X-Shooter spectra included in this work. References correspond to (Discovery, Redshift determination). The full list of references is given in the caption of Table~\ref{table:ESI}.} \label{table:Xarch}
\end{table}

\section{Data}\label{sec:observations}

\subsection{XQR-30}

\begin{figure*}
\includegraphics[width=\textwidth]{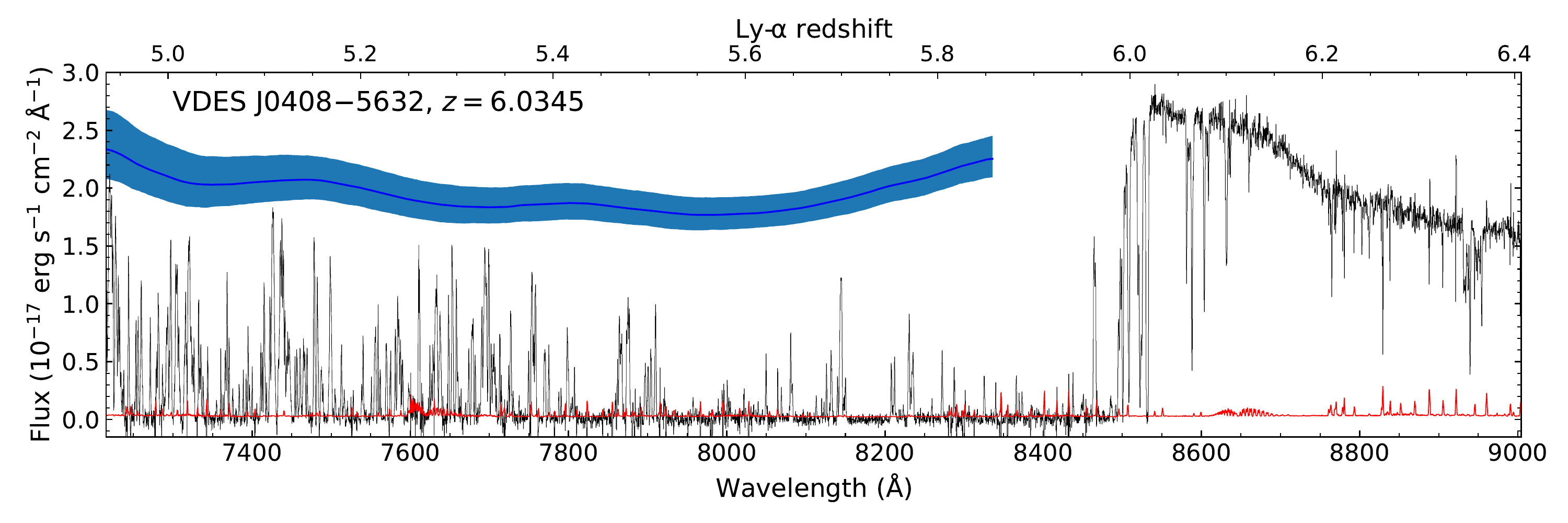}
\vspace{-2em}
\caption{X-Shooter spectrum of the \lal transmission region in the XQR-30 quasar VDES J0408-5632 at $z=6.0345$. The flux uncertainty is shown in red and the PCA-reconstructed continuum and its $\revmath{1\sigma}$ uncertainties are shown in blue. The PCA reconstruction is plotted over the wavelength range $1126\text{\AA}<\lambda<1185$\AA \ which we use in the mean flux measurement. The pixel scale is $10$ km s$^{-1}$ and the SNR of the \lal region \revi{(reconstruction divided by uncertainty)} is SNR $=86$. The exposure time was $13.5$ hours. } \label{fig:qso}
\end{figure*}

We primarily use data from the XQR-30 program (1103.A-0817(A)), which is ongoingly building a legacy sample of high-resolution spectra of $30$ quasars at $z\gtrsim5.8$ with the X-Shooter instrument \citep{XSHOOTER} on the Very Large Telescope. 
An example spectrum from the program is shown in Figure~\ref{fig:qso}. 
The XQR-30 quasars were selected to have the highest apparent luminosities at $z>5.7$. Observations were carried out using the $0.9''$ and $0.6''$ slits in the visible and near-infrared arms of X-Shooter, respectively. 
We use $25$ quasars from the XQR-30 sample which do not show strong broad absorption lines (BALs) precluding the modelling of the intrinsic continuum. We however retain the BAL quasars \hbox{ATLAS} J2211-3206, PSO J239-07 and PSO J239-07, whose BAL features are well-resolved and confined to highly-ionised absorption (Bischetti et al., in prep). All XQR-30 spectra have signal-to-noise ratios (SNRs) larger than $20$ per $10$ km s$^{-1}$ pixel measured over $1165\text{\AA}<\lambda<1170$\AA \ (Table~\ref{table:XQR30}). We use the reconstructed continua described in Section \ref{sec:reconstruction} to calculate the SNR over the range most relevant to our study. The X-Shooter instrument has a resolution of $\sim 34$ km s$^{-1}$ in the visible ($5500\text{\AA}<\lambda<10200$\AA) and $\sim 37$ km s$^{-1}$ in the infrared ($10200\text{\AA}<\lambda<24800$\AA), although better-than-average seeing during observations means the effective resolution is slightly higher. Observations are first flat-fielded and sky-subtracted following the method of \citet{Kelson03}, then the spectra extracted optimally \citep{Horne86} separately for the visible and infrared arms of the instrument. Our reductions routines are described in more detail in \citet{Becker09}; further details, including comparisons with the publicly-available \textit{esorex} \citep{esoreflex} and \textit{PypeIt} \citep{Pypeit-official} pipelines for X-Shooter, will be presented in D'Odorico et al.~(in prep). The optical and infrared arms are then stitched together over the $10110\text{\AA}<\lambda_\text{obs}<10130$\AA \ spectral window, by rescaling the infrared spectrum to match the observed mean flux in the optical arm after two rounds of sigma-clipping and discarding of all pixels with SNR$<2$. The spectrum is then interpolated over the overlap window. This somewhat aggressive procedure is adopted to minimize the risk of creating an artificial `step' in the spectrum between the arms, to which the continuum-fitting method may be non-linearly sensitive (c.f.~ \S \ref{sec:reconstruction}).

\subsection{Other X-Shooter spectra}
We supplement the XQR-30 quasars with $\revmath{26}$ archival X-Shooter spectra of equal SNR$>10$ per $10$ km s$^{-1}$ pixel from the literature (Table \ref{table:Xarch}), including three quasars at $5.5<z<5.7$ to better sample the \lal transmission at $z\leq 5.3$. The spectra were reduced in an identical manner to the XQR-30 quasars, except $6$ of them which had already been reduced with \textit{PypeIt}. \textit{PypeIt} is an open-source Python package designed to automate the reduction of spectroscopic data for (currently) $28$ different spectrographs \citep{pypeit}. Similarly to our custom reduction pipeline, \textit{PypeIt} performs joint extraction of objects and a model of sky emission in each observed frame. We conducted a comparative analysis on a sub-sample of quasars reduced via both methods, which showed only a negligible ($<0.5\%$) effect on the large-scale \lal transmission.


\subsection{ESI spectra}
Finally, we also complement our sample with $\revmath{16}$ archival spectra of $z\gtrsim 5.7$ quasars taken by the ESI instrument \citep{ESI} on the Keck Telescope (Table 3). The spectral resolution of ESI is lower than that of X-Shooter, at $\revmath{\sim60}$ km s$^{-1}$, and ESI's wavelength coverage only includes the optical up to $\lambda<10500$\AA. While we strive to reduce systematics arising from instrument and data reduction by minimising the number of different instruments, we include ESI spectra with SNR $>10$ per $15$ km s$^{-1}$ pixel since they constitute the largest collection of deep, publicly available observations of $z>5.7$ quasar spectroscopy with a single spectrograph besides X-Shooter. 
The ESI spectra were reduced using the same methods and algorithms as our X-Shooter pipeline, applying optimal spectral extraction after flat-fielding and sky subtraction. All but three of the ESI spectra we employ were also included in the `GOLD' sample of \citet{Bosman18}, where their reduction is further described. The three new spectra were reduced in an identical manner, but were not included in \citet{Bosman18} due to the availability of deeper MMTRCS \citep{MMTRCS} or HIRES spectroscopy \citep{HIRES}. Here we prefer the slightly shallower ESI spectra in order to preserve \revi{instrumental} consistency and reduce possible \revi{instrumentation} systematics. 
In a preliminary study \citepalias{Bosman21}, we analysed the impact of ESI's lesser resolution and wavelength coverage on systematics arising from quasar continuum reconstruction in the context of \lal transmission. We found that while continuum uncertainties were increased by $\sim50\%$ compared to using spectra with X-Shooter's wavelength coverage, no systematic biases arose. Six of our X-Shooter spectra were also observed to SNR $>10$ depth by ESI, enabling an empirical test of potential biases linked to instrumentation which we present in \S \ref{sec:instruments}.



\begin{table}
\begin{tabular}{l c c c}
Quasar ID & $z_{\text{qso}}$ & SNR pix$^{-1}$ & Refs. \\
\hline
\hline
SDSS J1148+5251 & $ 6.4189 $ & $ 118.8 $ & (33,34) \\
CFHQS J0050+3445 & $ 6.251 $ & $ 28.6 $ & (25,35) \\
SDSS J1623+3112 & $ 6.254 $ & $ 16.4 $ & (36,35)\\
SDSS J1250+3130 & $ 6.138 $ & $ 41.2 $ & (19,35) \\
SDSS J2315-0023 & $ 6.124 $ & $ 14.6 $ & (37,13) \\
SDSS J1602+4228 & $ 6.083 $ & $ 24.1 $ & (36,35) \\
SDSS J1630+4012 & $ 6.066 $ & $ 10.3 $ & (33,35) \\
SDSS J0353+0104 & $ 6.057 $ & $ 15.4 $ & (37,35) \\
SDSS J2054-0005 & $ 6.0389 $ & $ 22.6 $ & (37,27)  \\
SDSS J1137+3549 & $ 6.009 $ & $ 23.2 $ & (19,35) \\
SDSS J1411+1217 & $ 5.904 $ & $ 42.1 $ & (36,30) \\
\revi{SDSS J1335+3533} & $\revmath{ 5.9012 }$ & $\revmath{ 10.3 }$ &\revi{ (19,26) }\\
SDSS J0005-0006 & $ 5.847 $ & $ 18.4 $ & (36,13) \\
SDSS J0840+5624 & $ 5.8441 $ & $ 34.9 $ & (19,26) \\
SDSS J0002+2550 & $ 5.818 $ & $ 119.0 $ & (36,35) \\
SDSS J1044-0125 & $ 5.7846 $ & $ 64.9 $ & (38,27) \\
\hline
\end{tabular}
\caption{Quasars with archival ESI spectra included in this work. References relate to (Discovery, Redshift determination). (--) This paper; (1) \citet{Mazzucchelli17}; (2) D'Odorico et al.~in prep; (3) \citet{Reed17}; (4) \citet{Decarli18}; (5) \citet{Wang19}; (6) \citet{Chehade18}; (7) \citet{Farina19}; (8) \citet{Banados16}; (9) \citet{Wang16}; (10) \citet{Eilers20}; (11) \citet{DeRosa11}; (12) \citet{Jiang15}; (13) \citet{Becker19}; (14) \citet{Carnall15}; (15) \citet{Jiang16}; (16) \citet{Wang13}; (17) \citet{Banados14}; (18) \citet{Venemans15}; (19) \citet{Fan06}; (20) \citet{Wu15}; (21) \citet{Fan01}; (22) \citet{Jiang07}; (23) \citet{Mortlock09}; (24) \citet{Willott07}; (25) \citet{Willott10}; (26) \citet{Wang10}; (27) \citet{Venemans20}; (28) \citet{Venemans18}; (29) \citet{Schindler20}; (30) \citet{Kurk07}; (31) \citet{Morganson12}; (32) \citet{Yang17-qsos}; (33) \citet{Fan03}; (34) \citet{Willott15}; (35) \citet{Shen19}; (36) \citet{Fan04}; (37) \citet{Jiang08}; (38) \citet{Fan00}; (39) \citet{Wang21}; (40) \citet{Eilers20E}. 
For quasars without discovery papers, we reference the first paper which showcased or used a spectrum featuring broad emission lines. } \label{table:ESI}
\end{table}

\section{Methods}\label{sec:methods}

The effective \lal optical depth $\tau_\text{eff}$ is defined using the mean transmitted flux fraction in the \lal forest,
\begin{equation}
\tau_\text{eff} = -  \revmath{\ln} \left< \frac{F(\lambda)}{F_\text{cont}(\lambda)} \right>,
\end{equation}
where $F$ is the observed flux, $F_\text{cont}$ is the reconstructed intrinsic quasar continuum and $\langle \rangle$ is the mean over a fixed interval, traditionally taken to be $50$ cMpc/h (see \S \ref{sec:intervals}). 
The usable range of observed wavelengths is limited by the quasar's effect on its environment on one hand and overlap with \lab absorption on the other. To exclude the effect of the background quasars (the so-called `proximity zone', \citealt{Cen00,Carilli10,Eilers17}) we restrict ourselves to $\lambda<1185$\AA, beyond which no effect on \lal transmission is seen even in the deepest spectral stacks ($<0.5\%$ \lal flux increase: \citealt{Bosman18}). No quasars are known to have proximity zones extending beyond $1185$\AA: the longest $z>5$ proximity zone, in quasar SDSS J0100+2802, only extends to $\gtrsim 1189$\AA. In fact, we note that our proximity zone cut may be overly conservative, since no effect is seen in deep stacks even at $\lambda<1195$\AA \ at $z>6.1$ and the more conservative cut reduces the probed volume at $z>6.0$ by $\sim30\%$ for our sample. 

\begin{figure}
\includegraphics[width=\columnwidth]{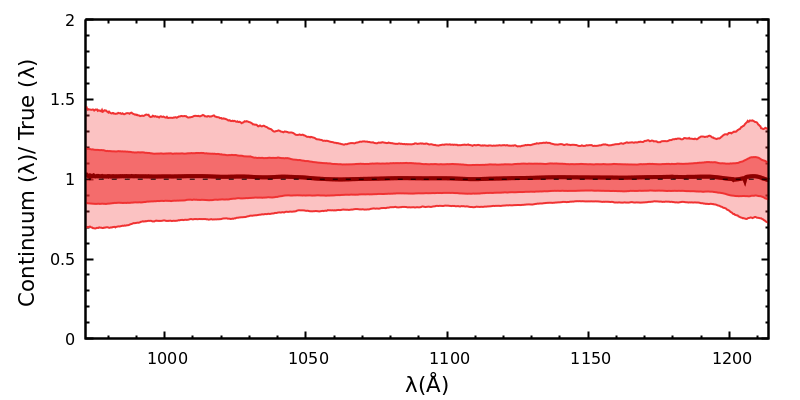}
\caption{Residuals in the PCA reconstruction of the $\lambda<1220$\AA \ blue-side continua of $4597$ eBOSS quasars at $2.7<z<3.5$, which were not used for training the PCA. No significant wavelength-dependent biases are seen. The average uncertainty over the $1026<\lambda<1185$\AA \ range, used in this paper, is $-7.9/+7.8\%$.} \label{fig:PCA}
\end{figure}

To exclude contamination by the overlapping \lab forest at low wavelengths, the redshift of the background quasar must be known precisely to determine its location with respect to the foreground IGM. When possible, we adopt the systemic redshifts of the quasar host galaxies, determined through the identification of sub-mm emission lines (these redshifts can roughly be identified in Tables~\ref{table:XQR30}, \ref{table:Xarch} and \ref{table:ESI} by having five significant digits). Redshifts may also be obtained from our rest-UV spectra directly using the quasar broad emission lines, but these features are often blue-shifted from the quasar host galaxies and from each other \citep{Meyer19-qso,Schindler20, Onoue20} with a large scatter $\sim 750$ km s$^{-1}$. An alternative method, which we use here, is to adopt the redshift of the first \lal absorber in front of the quasar \citep{Worseck14}. This method for locating the onset of the IGM has been shown to have relatively little offsets and scatter with respect to sub-mm emission lines, $\Delta v= 180\pm 180$ km s$^{-1}$ \citep{Becker21}. We employ it here for cases where fits to the Mg~{\small{II}} broad emission line are complicated by absorption, as indicated in Tables~\ref{table:XQR30}, \ref{table:Xarch} and \ref{table:ESI}.

To err on the side of caution, we round up the \lab wavelength of $1025.7$\AA \ and only use wavelengths $\lambda>1026$\AA. While the presence of the O~{\small{VI}} broad emission line renders the continuum prediction slightly more uncertain over the $1026\text{\AA}<\lambda\lesssim 1050$\AA \ wavelength range, this is carefully quantified and propagated to all our measurements and model comparisons (see Figure~\ref{fig:PCA} and \S \ref{sec:reconstruction}).  
We note that even if we use on occasion wavelengths contaminated by \lab absorption due to chance redshift errors, the corresponding \lb-absorbing gas would be located inside the quasar's proximity zone, and the \lab absorption should therefore be relatively small (although difficult to quantify in a model-independent manner).

The data reduction procedure in principle automatically rejects outlier pixels (e.g.~cosmic rays) when a large number of exposures are stacked. Nevertheless, we exclude a few ($\lesssim 0.05\%$ of total) anomalous pixels which are flagged if their SNR at the unabsorbed continuum level is $<2$ per pixel (since an average SNR$\geq 10$ is enforced for all our observations) or if pixels have negative flux at $>3\sigma$ significance. Such sigma-clipping can by definition only induce a bias $<\!\!<0.1\%$, while it cleans up features which are clearly reduction glitches.


\subsection{Redshift or length intervals?}\label{sec:intervals}

The traditional way of quantifying \lal optical depth fluctuations, motivated by efficiency when dealing with small sample sizes and by ease of comparison to theoretical models, has been to divide \lal transmission spectra in intervals of constant length \citep{Becker15, Bosman18, Eilers19, Yang20}. In this approach, the average transmission beyond a quasar's proximity zone is calculated over consecutive bins of fixed length (usually $\Delta L = 50$ cMpc/h) \revi{with variable starting and ending points, and these measurements are then assigned to a redshift interval depending on the mid-point of each bin.} We reproduce this approach for the purposes of comparison with the literature, but in our fiducial results we modify it for the purposes of comparison with theoretical models for two main reasons. First, the fixed length definition makes it possible for the same quasar sightline to contribute to the optical depth distribution in a single redshift bin more than once. This is a source of unwanted covariance, since \revi{the} IGM optical depth is known to be correlated on scales up to $100$ cMpc/h \citep{Becker15}. Second, the definition implies that up to half of \revi{the} pixels contributing to an optical depth measurement at a given redshift may be located outside of the redshift bin's bounds. The result is artificial scatter in measured optical depth, especially since \lal optical depth evolves very quickly at $z>5$ (\citetalias{Bosman21}; see also \citealt{Worseck16}). To circumvent these issues, we instead \revi{directly measure the opacity in bins with fixed starting and ending points of constant length in redshift space}. We divide the spectra in bins of $\Delta z= 0.1$ \revi{centred} at $z=5.0, 5.1, ..., 6.0$ \revi{corresponding to comoving lengths of $\revmath{l=36.0, 35.2, ... , 29.3}$ cMpc/h. We} retain measurements if $>50\%$ of the corresponding wavelength range is usable. In practice, $\sim 30\%$ of sightlines are truncated by more than $10\%$; we propagate the resulting uncertainties throughout the analysis.

For the purposes of future comparisons of the data with models, sometimes binning in equal length intervals, with fixed endpoints in redshift, might be preferable. We show the resulting distributions for $\Delta L = 50$ cMpc/h in Appendix \ref{app:length_binning}. Full measurements for $\Delta L = 30, 50, 100$ cMpc/h and $\Delta z = 0.05, 0.1$ are also available as online material.


\begin{figure}
\includegraphics[width=\columnwidth]{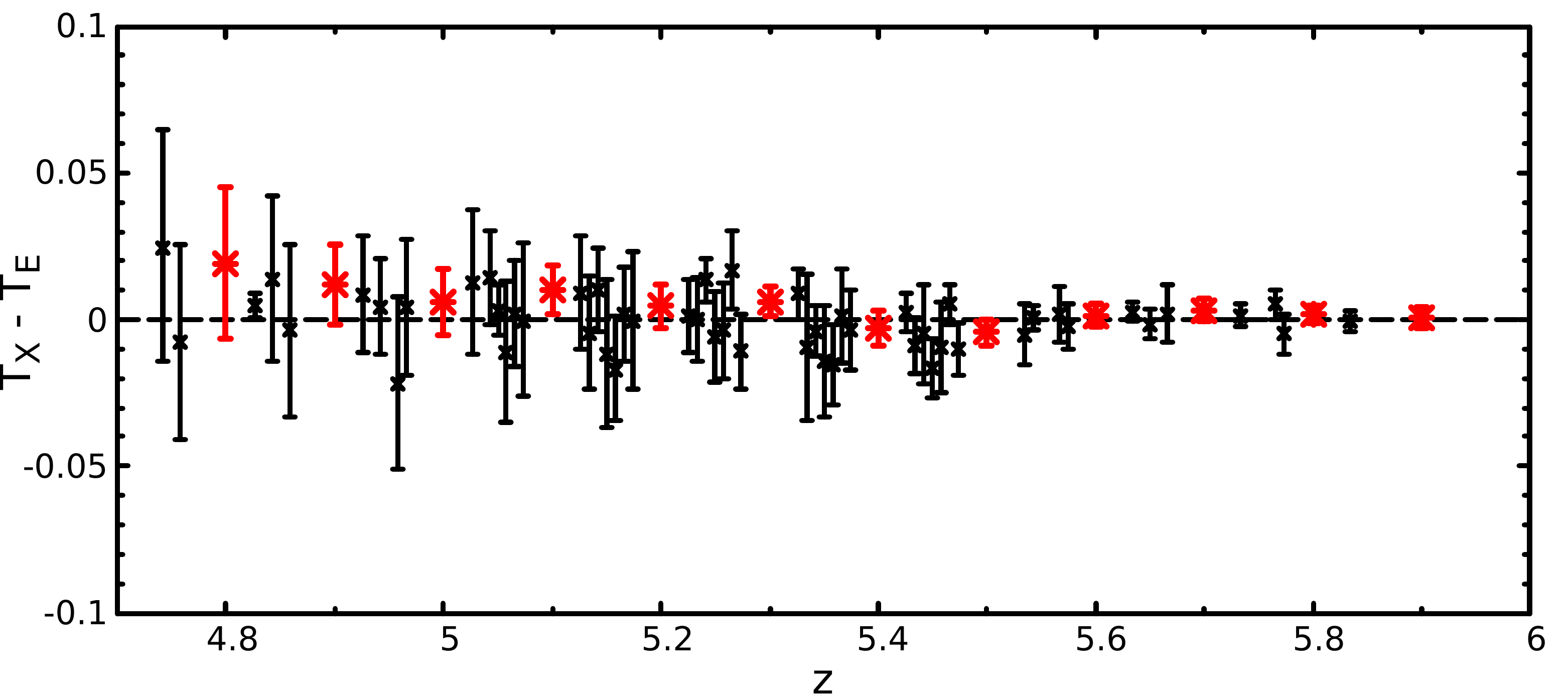}
\includegraphics[width=\columnwidth]{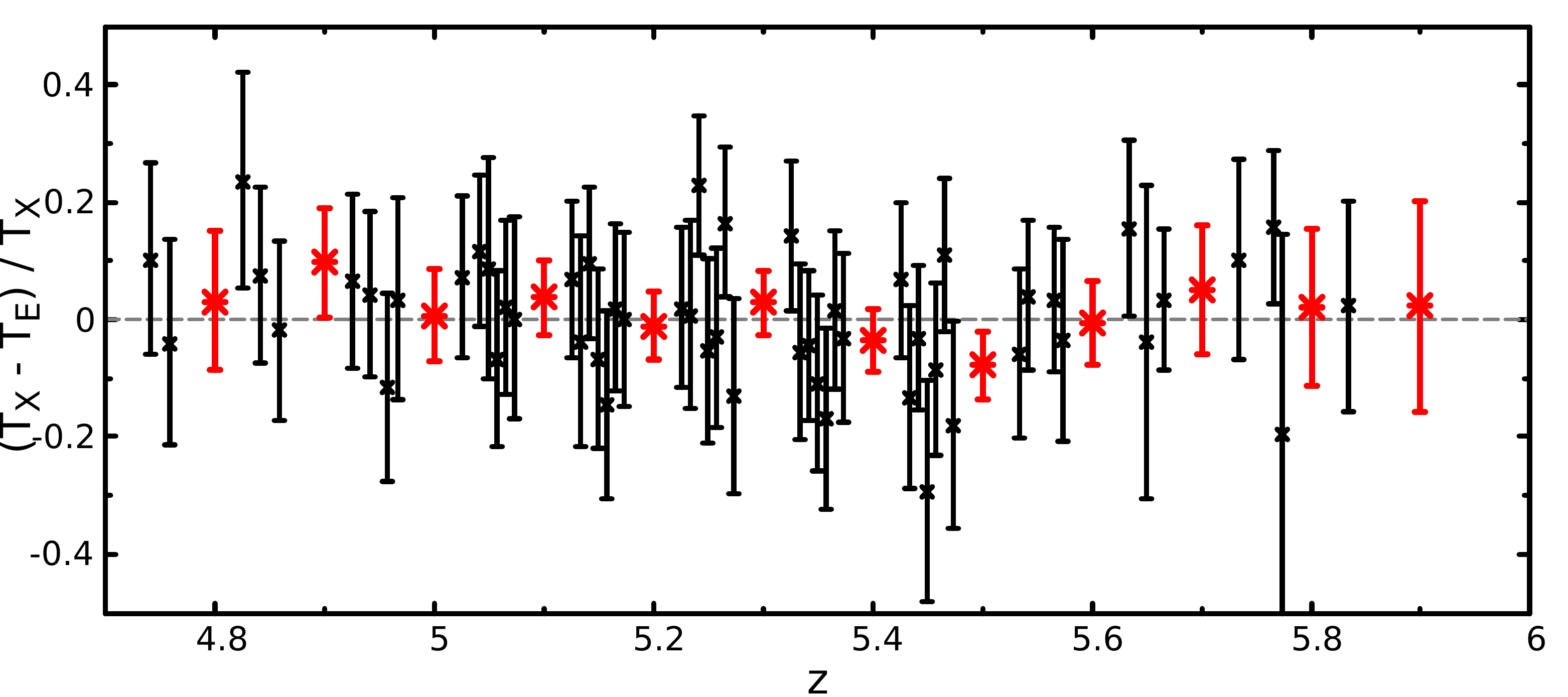}
\caption{Differences in \lal transmission measured with ESI ($T_E$) and X-Shooter ($T_X$) spectra of the same six quasars. \textit{Top:} Absolute difference in transmitted flux. The error bars account for observational uncertainties as well as continuum reconstruction uncertainties and biases. Individual measurements are shown in black with the averages in red. \textit{Bottom:} Same as top panel, but showing the fractional differences. The diagnostics show no evidence for instrument biases beyond the ones already accounted for in the measurement uncertainties. In both plots, some random scatter along the $x$-axis (redshift) has been added to improve legibility.}  \label{fig:inst}
\end{figure}

\subsection{Continuum reconstruction}\label{sec:reconstruction}

We employ Principal Component Analysis (PCA) to reconstruct $F_\text{cont} (\lambda)$ based on the observed quasar continuum at $\lambda>1280$\AA. Quasar continuum PCA models use a training set of low-$z$ quasar spectra to find optimal linear decompositions of the `known' red side ($\lambda>1280$\AA) and the `unknown' blue side of the spectrum ($\lambda<1220$\AA), then determines an optimal mapping between the linear coefficients of the two sides' decompositions \citep{Francis92, Yip04, Suzuki05, McDonald05, Paris11, Durovcikova20}. In \citetalias{Bosman21}, we conducted a rigorous comparison of the precision and accuracy of six reconstruction techniques used in the literature by using a large sample of `blind' tests with spectra where the true continuum was known. We found that two PCA methods outperformed both the more traditionally-employed power-law extrapolation (e.g.~\citealt{Bosman18}) and `stacking of neighbours' methods, both in prediction accuracy and in lack of wavelength-dependent reconstruction residuals. Here, we use a further improved version of the most accurate PCA method identified in \citepalias{Bosman21}, the log-PCA approach of \citet{Davies18-PCA} (see also \citealt{Davies18-DW}). 

Our PCA consists of $15$ red-side components and $10$ blue-side components. Training was performed on $4597$ quasars at $2.7<z<3.5$ with SNR$>7$ from the SDSS-III Baryon Oscillation Spectroscopic Survey (BOSS, \citealt{BOSS}) and the SDSS-IV Extended BOSS (eBOSS, \citealt{eBOSS}). Intrinsic continua were obtained automatically using a modified version of the method of \citet{Dallaglio08}, originally based on the procedures outlined in \citet{Young79} and \citet{Carswell82}.
The automatically-fitted continua are re-normalised to ensure they match the observed mean \lal transmission at $z\sim3$ measured from high-resolution spectra \citep{Faucher08, Becker13}, as they would otherwise be biased by the \revi{low} spectral resolution of the SDSS spectrograph (see discussion in \citealt{Dallaglio09}). 

Testing is performed by using an independent set of $4597$ quasars from eBOSS. 
The \revi{asymmetric} $1\sigma$ and $2\sigma$ bounds are measured by finding the central $68$th and $95$th percentile intervals of the prediction error in the testing sample at each wavelength. 
Figure~\ref{fig:PCA} shows the wavelength-dependent $1\sigma$ and $2\sigma$ continuum reconstruction uncertainties, Continuum$(\lambda)/$True$(\lambda)$. No features are visible at any rest-frame wavelength in the residuals, indicating that blue-side emission lines can be reconstructed without bias. The standard deviation is PCA$/$True $ - 1= 0.8_{-7.9}^{+7.8} \%$, i.e.~the method predicts the underlying continuum within $8\%$: a large improvement compared to power-law extrapolation methods ($>13\%$) and a slight improvement over the best PCA in \citetalias{Bosman21} ($9\%$). For the ESI spectra covering a shorter red-side wavelength range, we use the `optical-only' PCA developed in \citetalias{Bosman21} with PCA$/$True $ - 1= 1.0\%_{-11.3}^{+11.2}$. The lower accuracy is unsurprising since fewer features are available to the PCA modelling. However, no significant wavelength-dependent biases are present.

\begin{figure*}
\includegraphics[width=\textwidth]{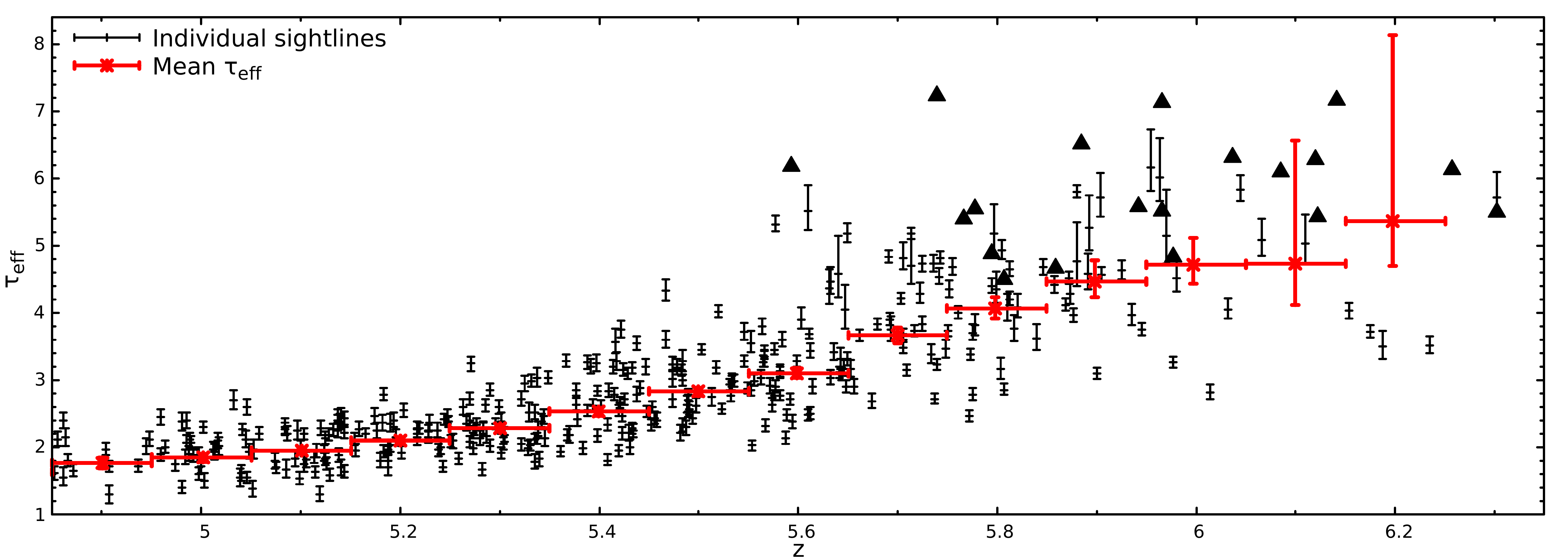}
\vskip-0.5em
\caption{Mean \lal flux measured along $67$ quasar sightlines at $4.9<z<6.2$, measured in \revi{consecutive} $50$ cMpc/h bins \revi{along each sightline (black)}. Non-detections are shown with upwards pointing triangles. The mean fluxes measured in intervals of $\Delta z = 0.1$ are shown with red points. Uncertainties correspond to the $16$th and $84$th percentile contours of a bootstrap resampling in each redshift interval. Non-detections are shown at the $2\sigma$ limit. The observed scatter between sightlines increases drastically above $z\gtrsim 5.4$.}\label{fig:large}
\end{figure*}

In the rest of the paper, we always correct for the residual wavelength-dependent mean bias ($<1\%$) 
to our reconstructions of $F_\text{cont}(\lambda)$ and forward-model the full wavelength-dependent uncertainties into all measurements and model comparisons. We refer the reader to \citetalias{Bosman21} for further details of the PCA training and testing procedures. Figures showing all PCA fits and blue-side predictions are shown in Zhu et al.~(submitted) and the PCA fits for XQR-30 spectra will be made public with the first XQR-30 data release (D'Odorico et al.~in prep).

\subsection{Instrumental effects}\label{sec:instruments}

To empirically check whether our data reduction and continuum reconstruction methods have accounted for all differences between ESI and X-Shooter spectra, we compare optical depth measurements for six quasars which have deep spectra with both X-Shooter and ESI. Figure~\ref{fig:inst} shows the difference between the \lal transmission observed with X-Shooter, $T_X = F_\text{X-Shooter} / F_\text{cont, X-Shooter}$ and with ESI, $T_E = F_\text{ESI} / F_\text{cont, ESI}$. The continua were reconstructed using the two PCAs discussed in \S \ref{sec:reconstruction}. The top and bottom panels show the absolute and fractional difference between $T_X$ and $T_E$, respectively. No statistically significant bias is detected at any redshift. Across all observations, the average fractional bias between the instruments is $1.3\%$ with an observed scatter of $2.3\%$. Since the effect is very sub-dominant compared to continuum uncertainties ($\sim11\%$ for ESI spectra) and we did not detect a statistically significant bias, we disregard instrumental differences between ESI and X-Shooter spectra beyond what is already included in the reduction pipelines.


\subsection{DLA exclusion}

Damped \lal absorbers (DLAs), named after their prominent \lal damping wings, are intervening systems along quasar sightlines with hydrogen column densities $N_\text{HI} \geq 10^{20.3}$ cm$^{-2}$ \citep{Wolfe05, Rafelski12}. DLAs near quasars at $z\gtrsim6$ can completely absorb \lal transmission over intervals $\Delta v = 2000$ km s$^{-1}$, with significant suppression of the transmission over $\Delta v \gtrsim 5000$ km s$^{-1}$ \citep{Dodorico18, Banados19, Davies20-ghost}. Since reionisation models typically do not include the effect of DLAs, we strive to remove them from our nominal measurements. 

\begin{table}
\centering
\begin{tabular}{l l c}
$z$ & $\left< F_{\text{Ly-}\alpha} \right> \ \ \ -1\sigma \ \ \ \ \ +1\sigma $ & $N_\text{los}$ \\
\hline
 $4.8$ & $0.194 \ \ \ -0.015 \ \ \ +0.018$ &  $15$ \\
 $4.9$ & $0.171 \ \ \ -0.014 \ \ \ +0.014$ &  $17$ \\
 $5.0$ & $0.1581 \ -0.0089 \ +0.0082$ &  $37$ \\
 $5.1$ & $0.1428 \ -0.0054 \ +0.0068$ &  $48$ \\
 $5.2$ & $0.1222 \ -0.0054 \ +0.0046$ &  $55$ \\
 $5.3$ & $0.1031 \ -0.0050 \ +0.0056$ &  $58$ \\
 $5.4$ & $0.0801 \ -0.0048 \ +0.0061$ &  $64$ \\
 $5.5$ & $0.0591 \ -0.0035 \ +0.0039$ &  $64$ \\
 $5.6$ & $0.0447 \ -0.0036 \ +0.0033$ &  $59$ \\
 $5.7$ & $0.0256 \ -0.0029 \ +0.0031$ &  $51$ \\
 $5.8$ & $0.0172 \ -0.0028 \ +0.0022$ &  $45$ \\
 $5.9$ & $0.0114 \ -0.0030 \ +0.0029$ &  $28$ \\
 $6.0$ & $0.0089 \ -0.0029 \ +0.0033$ &  $19$ \\
 $6.1$ & $0.0088 \ -0.0074 \ +0.0082$ &  $10$ \\
 $6.2$ & $0.0047 \ -0.0044 \ +0.0045$ &  $8$  \\
\end{tabular}
\caption{Mean \lal flux transmission at $4.75<z<6.25$, measured in $\Delta z = 0.1$ intervals centred on the redshift given in the first column. Uncertainties correspond to the $16$th and $84$th percentiles from bootstrap resampling. The measurement uncertainties on their own are a factor $5-10$ smaller than the bootstrap uncertainties quoted here. $N_\text{los}$ sightlines contribute to each measurement.}\label{table:means}
\end{table}

The detection of $z\gtrsim5$ DLAs relies on the identification of associated low-ionisation metal absorption lines, since their \lal absorption may not contrast against the highly-opaque IGM. DLA metallicities at $z\gtrsim5$ are very diverse, and some can be highly sub-solar \citep{Banados19}, such that even relatively weak metal absorption might indicate a DLA. The identification of intervening metal absorbers in the XQR-30 sample will be described in detail in Davies et al.~(in prep). For the other quasars, we used where relevant the published lists of intervening metal systems of \citet{Cooper19}, \citet{Dodorico18}, \hbox{\citet{Meyer19} and \citet{Becker19}}. We supplemented the literature where necessary by conducting our own metal search, following closely the standard procedure described in \citet{Bosman17}. Pairs of absorption lines corresponding to the same ion or frequently co-occurring ions (C~{\small{IV}}, Mg~{\small{II}}, Fe~{\small{II}}, O~{\small{I}}+C~{\small{II}}) are searched for automatically before being confirmed manually. Due to the high SNR of the X-Shooter spectra, we expect to be $>90\%$ complete to absorption corresponding to $\log N_\text{Mg II}/$cm$^{-2}\gtrsim 13$. The metal identification in the ESI spectra similarly relies on literature studies which employed infrared spectra of the objects. 

We adopt the following criteria: we mask the central $\Delta v = 3000$ km s$^{-1}$ for systems with metal column densities $\log N_\text{C II}/$cm$^{-2}>13$, $\log N_\text{O I}/$cm$^{-2}>13$, or $\log N_\text{Si II}/$cm$^{-2}>12.5$, measured through the $\lambda=1334.53$\AA, $1302.16$\AA, and $1526$\AA\  transitions, respectively. When none of these ions are accessible, we also exclude the central $\Delta v = 3000$ km s$^{-1}$ for systems with $\log N_\text{Mg II}/$cm$^{-2}>13$ based on the high rates of co-occurrence of the Mg~{\small{II}} $2796.35, 2803.53$\AA \ doublet \citep{Cooper19}. We exclude a larger window of $\Delta v = 5000$ km s$^{-1}$ around intervening systems with $\log N_\text{O I, C II, Si II, Mg II}>14$ cm$^{-2}$ due to the likely presence of extended damping wings. 

We do not exclude systems based on the presence of highly-ionised ions alone (e.g.~C~{\small{IV}}, Si~{\small{IV}}) since the corresponding gas is likely highly ionised \citep{Cooper19}. Finally, we exclude $\Delta v = 5000$ km s$^{-1}$ around the suspected location of strong O~{\small{VI}} associated absorption (from systems detected from strong associated C~{\small{IV}} absorption), which overlaps with the \lal forest in quasars J1411+1217 and J1623+3112.



\section{Results} \label{sec:results}

Figure~\ref{fig:large} shows the mean \lal transmission measured in intervals of $50$ cMpc/h. The average transmission evolves smoothly over $5<z<6$, but an increase in scatter between measurements at equal redshift becomes clear at $z\geq5.4$. By $z=5.6$, the sampling of the distribution is visibly limited. The number of fully opaque Gunn-Peterson troughs with non-detections ($2\sigma$) at $\tau>6$ increases sharply, with the first occurrence found at $z\sim5.6$.

\subsection{Mean transmission across $4.8<z<6.2$}

\begin{figure}
\includegraphics[width=\columnwidth]{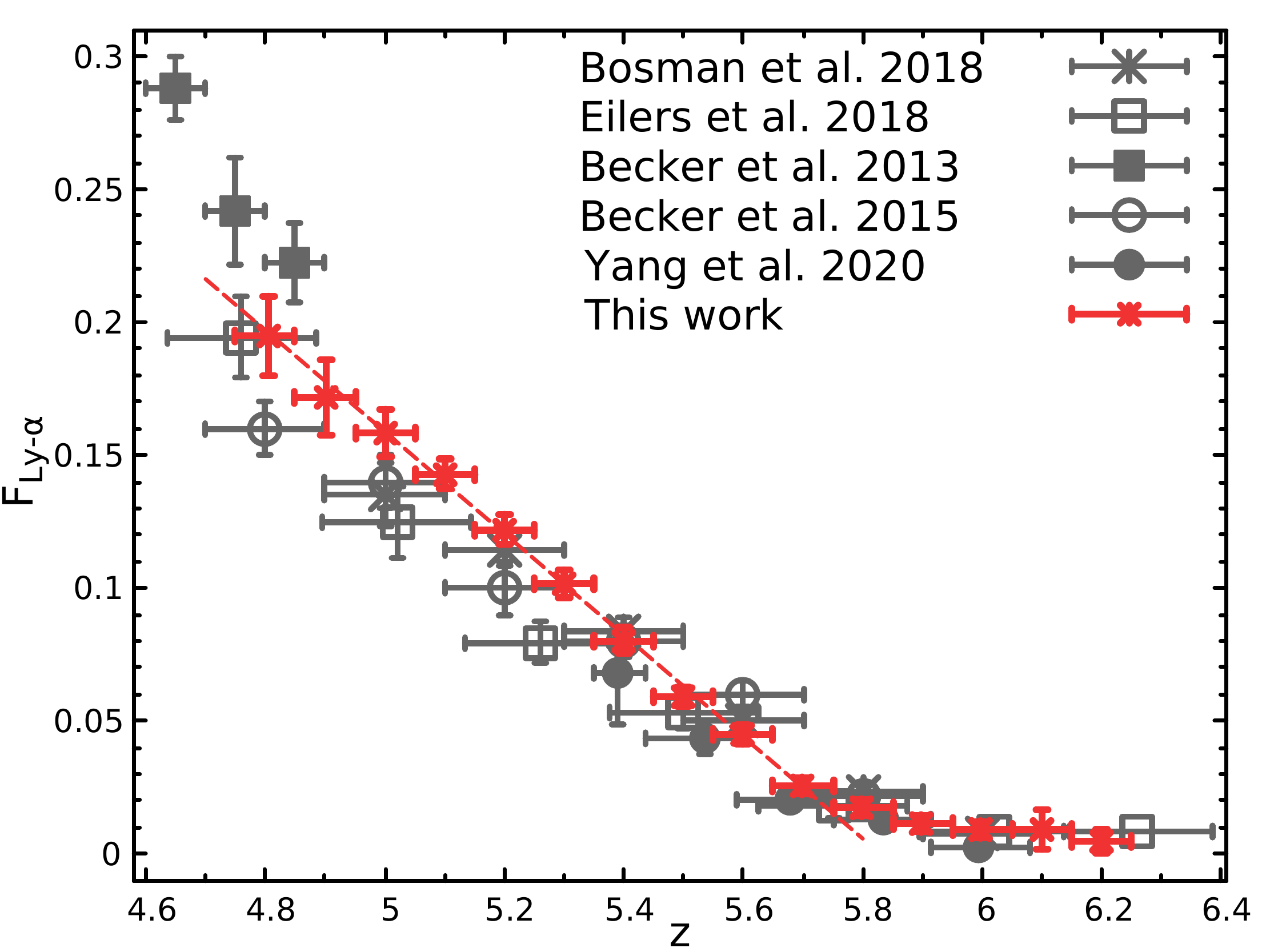}
\caption{Average \lal transmission evolution with redshift. Uncertainties are obtained via bootstrap resampling in this work as well as in the literature comparison samples of \citet{Becker13, Becker15, Bosman18, Eilers18}, and should therefore encompass cosmic variance as long as the underlying optical depth distributions are well-sampled. Differences at $z\gtrsim5.5$ are due to under-sampling of cosmic variance, as well as systematic biases in older work. At $z>5.4$, cosmic variance is $5-10$ times larger than measurement uncertainties. The red dashed line shows the optimal linear fit to the data over $4.8<z<5.7$ (Equation 2, see text).} \label{fig:means}
\end{figure}

\begin{figure}
\includegraphics[width=\columnwidth]{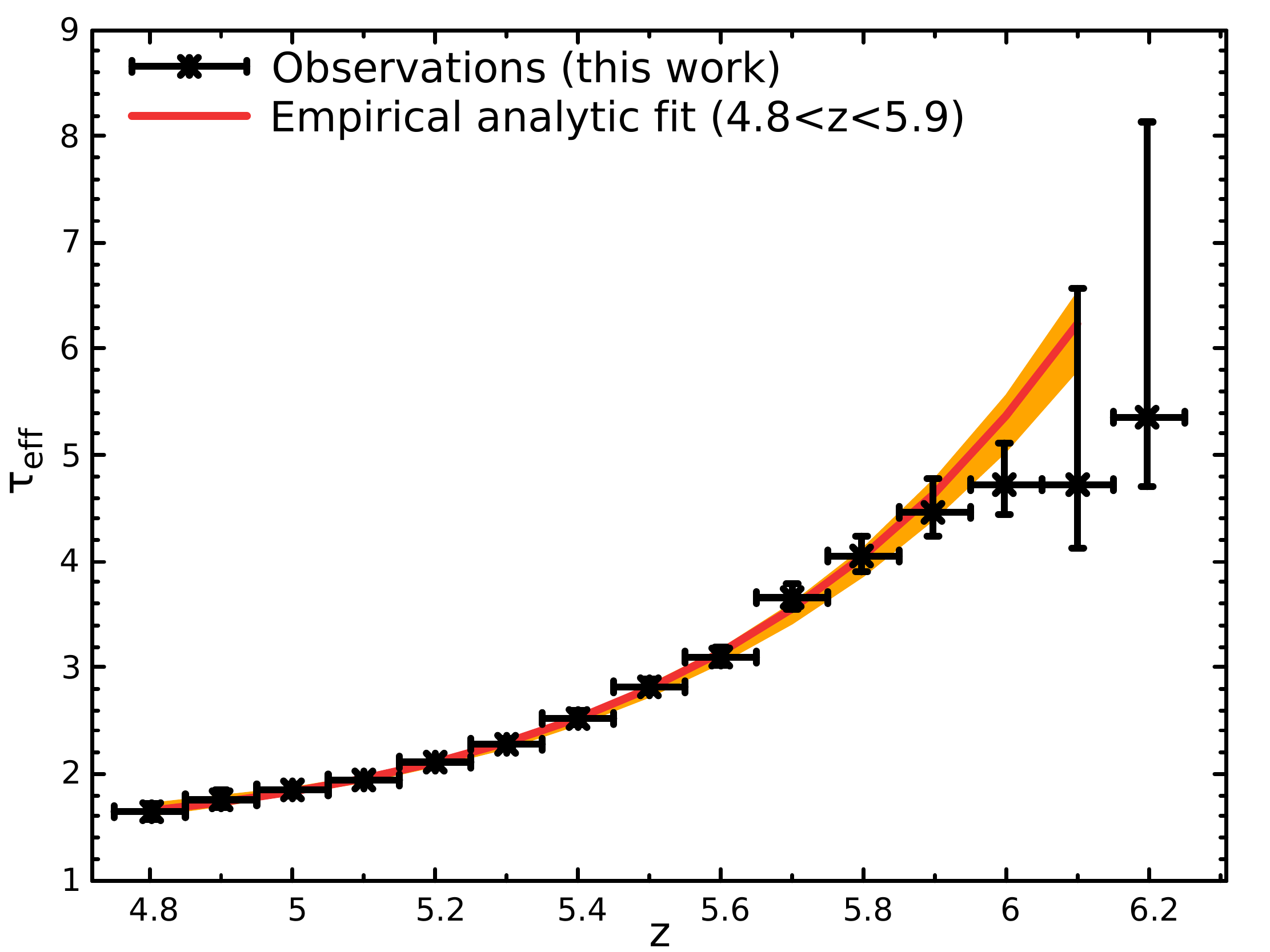}
\caption{Evolution of $\tau_\text{eff}$ with redshift measured in our sample across $4.8<z<6.2$ in steps of $\Delta z = 0.1$ (black). The red line shows the best-fit power-law model and its $68\%$ uncertainty envelope (Equation 3, see text). Observational uncertainties are obtained via bootstrap resampling. } \label{fig:means_tau}
\end{figure}

We calculate the mean transmission in bins of $\Delta z = 0.1$ and give the results in Table~\ref{table:means}. We do not weight the measurement: all pixels corresponding to \lal transmission inside a given redshift interval contribute equally (after the masking of bad regions as described above). The uncertainties are calculated via bootstrap re-sampling in each redshift bin. We quote the $16\%$ and $84\%$ percentiles of the bootstrap results. The observational uncertainties, taking into account only uncertainties in individual measurements, are a factor $5-10$ smaller than the bootstrap uncertainties at all redshifts. The uncertainties are therefore dominated by the intrinsic width of the \lal transmission distribution.  
The mean \lal transmission over the range $4.8<z<5.7$ is empirically well-described by a linear decline of the form:
\begin{equation}
F_{Ly-\alpha}(1+z) = a \times (1+z) + b.
\end{equation}
We fit this functional form to our observations using least-squares regression, and obtain best-fitting parameters $a=-0.191, b=1.307$. Both parameters are constrained to better than $0.1\%$. We show the resulting curve in Figure~\ref{fig:means}. An empirical parametric description of effective \lal optical depth evolution with redshift, used for instance by \citet{Becker13}, is an power-law function with a constant offset of the form:
\begin{equation}
\tau_\text{eff}(1+z) = \tau_0 \left(\frac{1+z}{1+z_0}\right)^\beta + C.
\end{equation}
Setting $z_0=4.8$, we run a least-squares regression and find best-fit parameters $\tau_0 = 0.30 \pm 0.08$, $\beta = 13.7 \pm 1.5$ and $C = 1.35\pm 0.12$. We fit this form to the mean optical depth over $4.8<z<5.9$ and show the resulting best-fit model in Figure~\ref{fig:means_tau}. 
We sample the covariance matrix of the three parameters and calculate the upper and lower envelopes encompassing $68\%$ of the variance about the best fit, which are shown by the orange shaded region. The evolution of $\tau_\text{eff}$ with redshift is much steeper at $z>4.8$ than over $2<z<5$, where \citet{Becker13} found a best-fit $\beta = 2.90$. 

Our measurements are in fair agreement with past literature, as shown in Figure~\ref{fig:means}. The quasars used in this work have considerable overlap ($\sim30-50\%$) with the ones employed by \citet{Becker15, Eilers18, Bosman18} and \citet{Yang20}, such that differences are unlikely to be due to cosmic variance alone. Systematic differences in continuum reconstruction methods are a known cause of bias: as shown in \citetalias{Bosman21}, the tension between the measurements of \citet{Bosman18} and \citet{Eilers18} can be explained almost entirely by the different continuum reconstruction methods employed \revi{by the two studies. \citet{Bosman18} employed power-law extrapolation, while \citet{Eilers18} used a linear PCA originating in a small number of hand-fitted continua in \citealt{Paris11}. 
Both methods were found to introduce non-trivial wavelength-dependent biases which are virtually absent from more recent log-space PCA and neural-network-mapped PCAs (e.g.~\citealt{Davies18-PCA, Durovcikova20}; see \citetalias{Bosman21} for details). Such biases depend sensitively on the redshifts of the background quasars and corrections unfortunately cannot be applied post-hoc.} 
\citet{Becker15}, \citet{Bosman18} and \citet{Yang20} all employed power-law reconstructions, and therefore carry similar biases; this may explain why our results are offset from all three studies in the same direction at $z<5.4$ (where power-law-induced uncertainties and biases are the largest). We also note that \citet{Eilers18}, \citet{Bosman18} and \citet{Becker15} had substantial overlap in quasar sightlines, and should therefore be affected by cosmic variance in a coherent way compared to our sample. \citet{Yang20} calculated mean optical depths by using a weighted spectral stack, without providing measurements of scatter between sightlines. In order to provide a better comparison with this study, we re-calculate the optical depth values from \citet{Yang20}'s sample by using their published list of measured optical depths in each quasar sightline, and we estimate the cosmic variance uncertainties via bootstrap resampling. 



\begin{figure*}
\includegraphics[width=\textwidth]{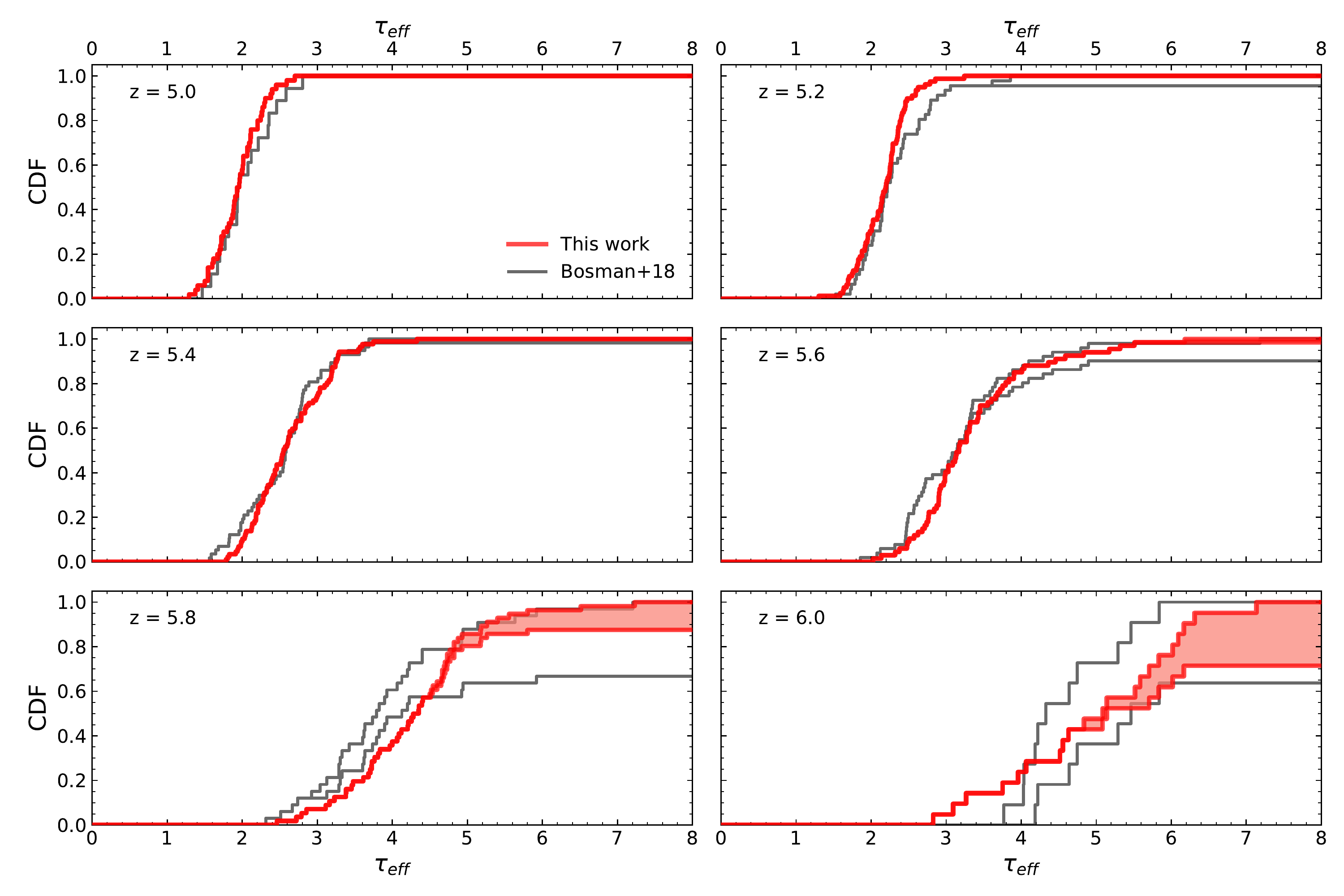}
\vskip-1em
\caption{CDFs of \lal optical depth (red) computed in $50$ cMpc/h intervals, compared to results from \citet{Bosman18} (black). The lower and upper CDFs correspond to assumptions that non-detections are infinitely opaque, or just below the detection threshold, respectively. While the sample sizes are comparable ($N=67$ and $N=64$, respectively), the highly improved data quality and new correction of systematic biases linked to instrumentation and continuum reconstruction result in overall smoother distributions and fewer non-detections. The most transparent sightlines at $z=6.0$ are shown in Figure~\ref{fig:rare}}.  \label{fig:compar}
\end{figure*}
\begin{figure}
\includegraphics[width=\columnwidth]{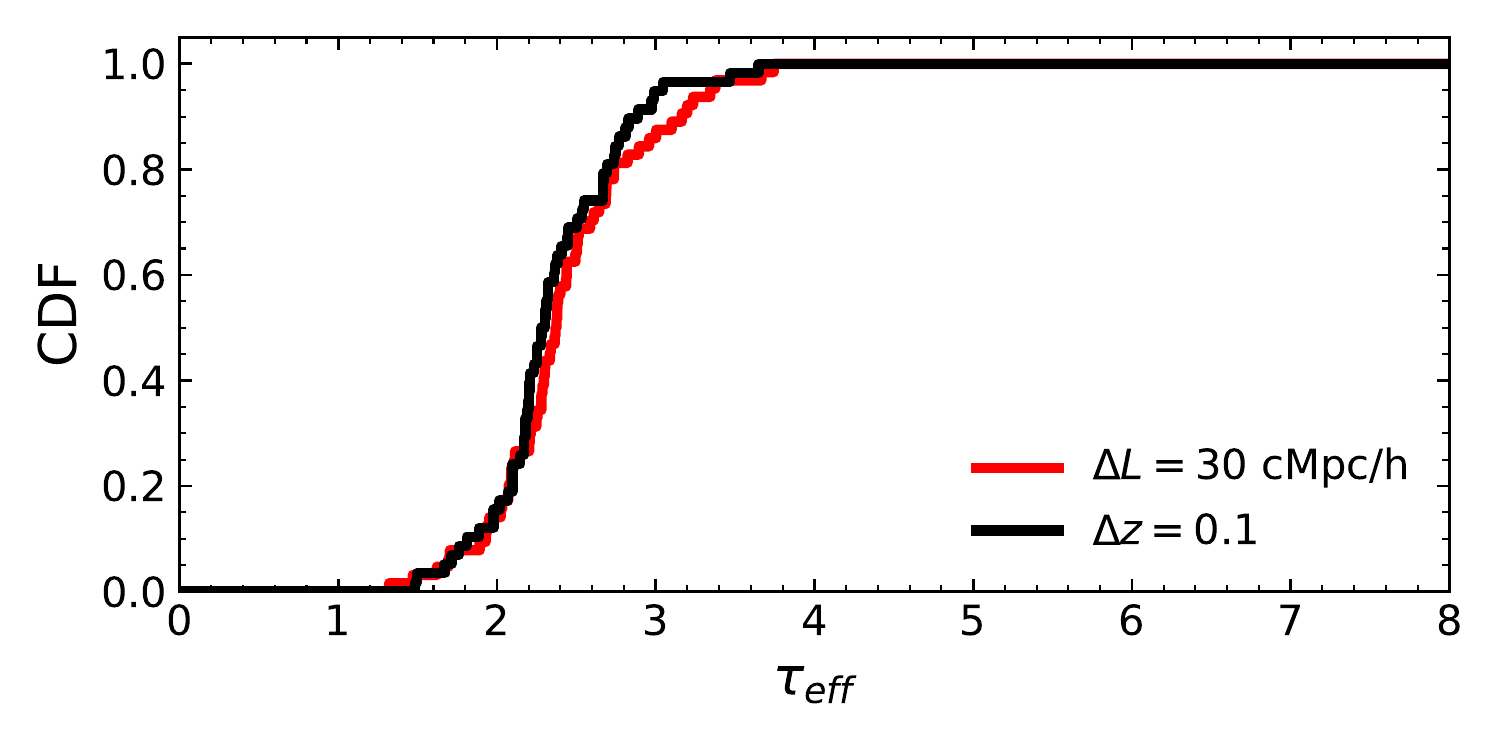}
\vskip-1em
\caption{Comparison of the optical depth distribution measured over the same interval, $5.25<z<5.35$, using bins of constant length with boundaries that vary (red) or constant $\Delta z$ boundaries (black). The length of $30$ cMpc/h corresponds to the redshift interval $\Delta z = 0.1$; the only difference between the distributions is the definition of binning and not the lengths over which the optical depth is intrinsically computed. The excess of highly opaque sightlines in the red curve originates from length bins whose centres lie near the top end of the range ($z\sim5.35$), artificially increasing the scatter by including contributions from pixels outside the nominal redshift range. To avoid this bias, we adopt binning in constant redshift for the purposes of inference.}  \label{fig:zbin}
\end{figure}

\begin{figure*}
\includegraphics[width=0.98\textwidth]{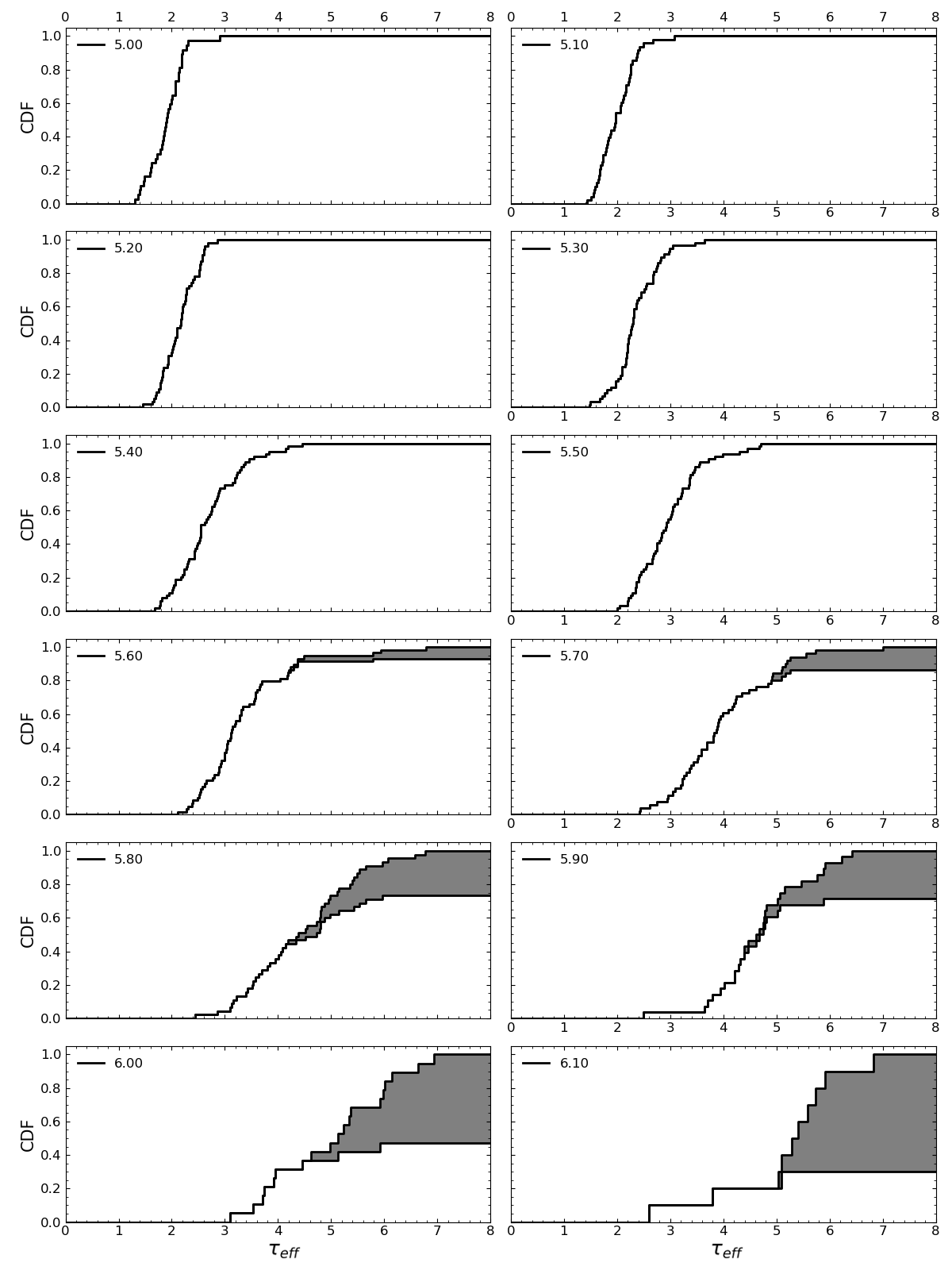}
\caption{CDFs of \lal optical depth in bins of constant $\Delta z = 0.1$. The distributions begin to appear `elongated' around $z\sim5.4$. The fraction of non-detections (as seen by the difference between the curves at the right edges of the CDFs) increases sharply above $z=5.7$, but some transparent sightlines remain even at $z=6.1$. The most transparent sightlines at $z=5.9$ and $z=6.1$ are shown in Figure~\ref{fig:rare}.}  \label{fig:dz}
\end{figure*}

\begin{figure*}
\includegraphics[width=\textwidth]{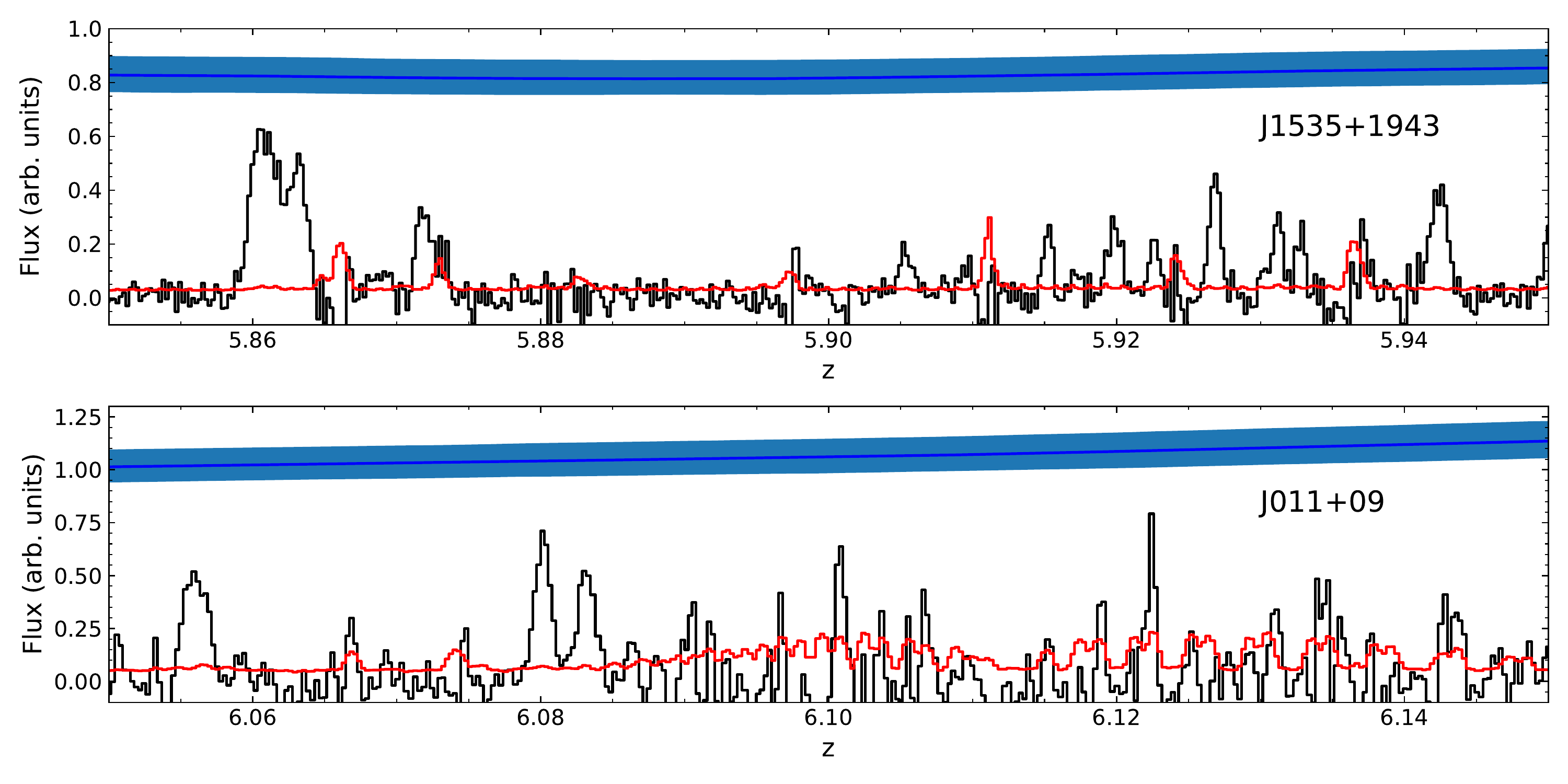}
\caption{Black: Transmitted \lal flux in the most transmissive sightlines at $z=5.9$ (J1535+1943, top) and at $z=6.1$ \revi{(PSO J011+09, bottom)}. Red: observational uncertainties. Blue: PCA continuum reconstruction and its $\pm1\sigma$ uncertainties. The $6.10<z<6.14$ range is affected by increased uncertainties due to sky emission corrections, but excluding this region slightly raises the \revi{overall} mean transmitted flux over the $6.05<z<6.15$ interval.}  \label{fig:rare}
\end{figure*}

We find a very smooth, linear evolution of the mean \lal transmitted flux across $4.8<z<5.7$, with no sudden steepening, in contrast with several past studies. Matching the results of \citet{Becker13} at $z<4.8$ still seems to require a faster steepening at $z\sim4.7$ (Fig.~\ref{fig:means}). However, both their measurements and ours are at the (opposite) edges of their redshifts of validity at $z\sim4.8$. In our study, measurements at $z<5$ rely on $N<20$ objects, and are the most sensitive to errors in background quasar redshift via contamination by \lab absorption. Conversely, the measurements of \citet{Becker13} at $z>4.5$ use the \lal forest at the shortest separations from the background quasars, where the continuum is under-predicted by power-law extrapolations due to the large width of the broad \lal emission line \citepalias{Bosman21}; the measurements of \citet{Becker13} are also based on spectra with much lower SNR than our sample. Properly sampling the overlapping region would therefore require a sample of deep quasar spectra of intermediate redshifts, i.e.~$5<z_\text{qso}<5.5$.



\subsection{Optical depth distributions at $5.0<z<6.1$}

We show the distributions of \lal optical depth at $5.0<z<6.1$ for comparison with previous studies in Figure~\ref{fig:compar}. We first use the traditional method of measuring the optical depth in constant $50$ cMpc/h intervals whose centres are then binned by redshift, since this definition was employed by all past studies. We will highlight the biases created by this definition later in Figure~\ref{fig:zbin}.

Non-detections of transmission over a given interval (defined at the $2\sigma$ level) give rise to lower limits on optical depth. Limits can either be represented as corresponding to flux equal to twice the measurement uncertainty (i.e.~just below the detection threshold, e.g.~\citealt{Becker15}) or as corresponding to infinite optical depth. Following \citet{Bosman18}, we display the cumulative distribution functions (CDFs) showing both bounds. The lower CDF assumes that all non-detections are infinitely opaque, while the upper CDF assumes all non-detections correspond to flux just below the detection limit.


Owing to the much higher SNR of our sample, the number of non-detections over $50$ cMpc/h is highly reduced at $5.5<z<5.9$ compared to \citet{Bosman18}, which used a sample of size comparable to ours ($N_\text{qso}=64$ compared to our $N_\text{qso}=67$) but with widely varying SNR. Our study only employs spectra which are able to probe optical depths up to (at least) $\tau=4.5$ in $\Delta z = 0.1$ bins. At $5.5<z<5.7$, the only non-detection is in quasar PSO J025-11 ($1/67$ sightlines) while $\sim10\%$ of sightlines were undetected in \citet{Bosman18}. 
The quasar J0148+0600 (the longest trough from \citealt{Becker15}) was formerly the most opaque at this redshift, but a slight update to the quasar's redshift shifts the  exact start and end of the measurement bin in our study such that transmission is detected ($\tau = 5.33$) in a bin centred at $z=5.577$. This issue highlights one of the problems with the classical definition of optical depth binning: results depend non-trivially on the assumed redshifts of the background quasars (while our binning \revi{explicitly} does not). At $5.7<z<5.9$, the number of non-detections is reduced from $\sim35\%$ (in \citealt{Bosman18}) to $12.5\%$ ($7/56$ sightlines have $\tau>4.5$). In contrast, our increased sensitivity does not reduce the number of sightlines with non-detections at $5.9<z<6.1$, where $\sim30\%$ of sightlines ($6/21$) remain fully absorbed. A large fraction of $z\gtrsim5.9$ sightlines are therefore more opaque than $\tau=4.5$, a limit which is unlikely to be exceeded for large samples of quasars with current instrumentation. Significant advances in sensitivity, which may be required to detect residual transmission in the bulk of quasars at $z>6$, could be brought by the next-generation Extremely Large Telescope \citep{ELT} or the Thirty Meter Telescope \citep{TMT}.


To determine the lowest redshift at which the optical depth distribution is in agreement with fluctuations from density alone, we use redshift bins with $\Delta z = 0.1$. We choose this binning size in order to resolve the fast evolution in the mean optical depth (Figure \ref{fig:means}). In Section 3.1 we highlighted some potential biases inherent to the classical definition of binning optical depth measurements of constant length. In Figure~\ref{fig:zbin} we demonstrate these biases, which become more pronounced as the redshift intervals are shortened. The optical depth distribution at $z=5.3$ is artificially broadened by the inclusion of transmission outside the nominal redshift range, as shown in the red curve. This effect is non-negligible when the binning length becomes comparable to the redshift interval, when the evolution in the mean is rapid (as shown in Figure~\ref{fig:means}), or when using sightlines near the end of their usable wavelength ranges (i.e.~the lowest redshift bins). We adopt binning in constant redshift intervals for the purposes of inference in order to avoid this bias, and show the resulting CDFs in $\Delta z=0.1$ in Figure~\ref{fig:dz}. The new definition also avoids any covariance of sightlines in distributions at fixed redshift. The distributions will still be covariant between redshifts, since opaque sightlines show coherence over scale $\Delta z>0.1$.


\subsection{$z\geq5.9$ transmissive sightlines}

In Figure~\ref{fig:rare} we show the most transmissive sightlines at $z=5.9$ and $z=6.1$ which stand out from the distributions in Figure~\ref{fig:dz}. The XQR-30 quasar J1535+1943 is the most transmissive at $z=5.9$ with $\tau_\text{eff}=2.50$, showing \lal transmission over the entire $5.85<z<5.95$ interval (top panel). The same quasar is the second most transmissive at $z=6.1$, with $2$ strong transmission spikes at $z\sim6.07$ resulting in $\tau_\text{eff}=3.79$; this suggests elevated transmission over scales $\gtrsim 100$ cMpc/h. At $z=6.1$, the X-Shooter archival quasar PSO J011+09 is the most transmissive by far, with $3$ very strong transmission spikes resulting in $\tau_\text{eff} = 2.59$ (bottom panel). The transmission is affected by increased uncertainties due to corrections for telluric absorption; however, excluding the regions affected \revi{by increased uncertainties} actually further lowers the measured optical depths. 
\revi{J1535+1943 and PSO J011+09 display flux transmission larger than the mean at $\revmath{z=5.9}$ and $\revmath{z=6.1}$ by factors of $\revmath{4.3}$ and $\revmath{4.9}$, respectively (corresponding to optical depths $\revmath{50\%}$ smaller than the mean).} 
The discovery of such rare transmissive sightlines is only possible by employing large samples of quasars to sample cosmic variance: at $z=5.9$, only $1/51$ sightlines has an optical depth $\tau_\text{eff}<3$. Characterising the extrema of the optical depth distribution at fixed redshift is crucial in order to design models of UVB fluctuations which reproduce the full variety of environments at the end of reionisation. The quasar J1535+1943 was not included in any previous measurements of optical depth; its addition to our sample raises the average transmitted flux by $\sim 10\%$. While this change is comfortably included within our quoted bootstrap uncertainties, it may account for some of the systematic disagreements between our study and past work which did not include this quasar (Figure~\ref{fig:means}).

In addition, \lal transmission spikes can been used to measure the thermal state of the IGM (e.g.~\citealt{Gaikwad20}) and to pose \revi{constraints} on reionisation history through their statistical distribution \citep{Barnett17,Chardin17-spikes}. The identification of strong transmission spikes at $z>5.8$ therefore opens up complementary analyses, which will explored in a separate paper (XQR-30 in prep).



\begin{table}
\centering
\begin{tabular}{l c c}
$z$ & $\Gamma_\text{Sherwood} /10^{-13} \text{s}^{-1}$ & $\Gamma_\text{Nyx} /10^{-13} \text{s}^{-1}$ \\
\hline
$5.0$ & $7.85$ & $7.34$ \\
$5.1$ & $7.57$ & $6.92$ \\
$5.2$ & $7.39$ & $6.62$\\
$5.3$ & $7.03$ & \phantom{$(6.13)$} \  $7.32$ \ $(6.13)$ \\
$5.4$ & $6.17$ & $6.41$\\
$5.5$ & $5.77$ & $5.56$\\
$5.6$ & $4.46$ & $4.47$ \\
$5.7$ & $4.05$ & $4.02$\\
$5.8$ & $3.79$ & $3.94$ \\
\hline
\end{tabular}
\caption{\revi{Rescaled \revi{photo-ionisation} rates which maximise the likelihood of the observed optical depth distributions for Sherwood and Nyx at each redshift. The rescaling factors themselves are in the range $0.5-2$. The discontinuity at $z=5.3$ in Nyx is to due switching from the $z=5.0$ to the $z=5.5$ snapshots (see text); the value in brackets gives the rescaled \revi{photo-ionisation} rate when the $z=5.0$ snapshot is used instead of $z=5.5$.}}\label{table:resc}
\end{table}

\begin{figure*}
\includegraphics[width=\textwidth]{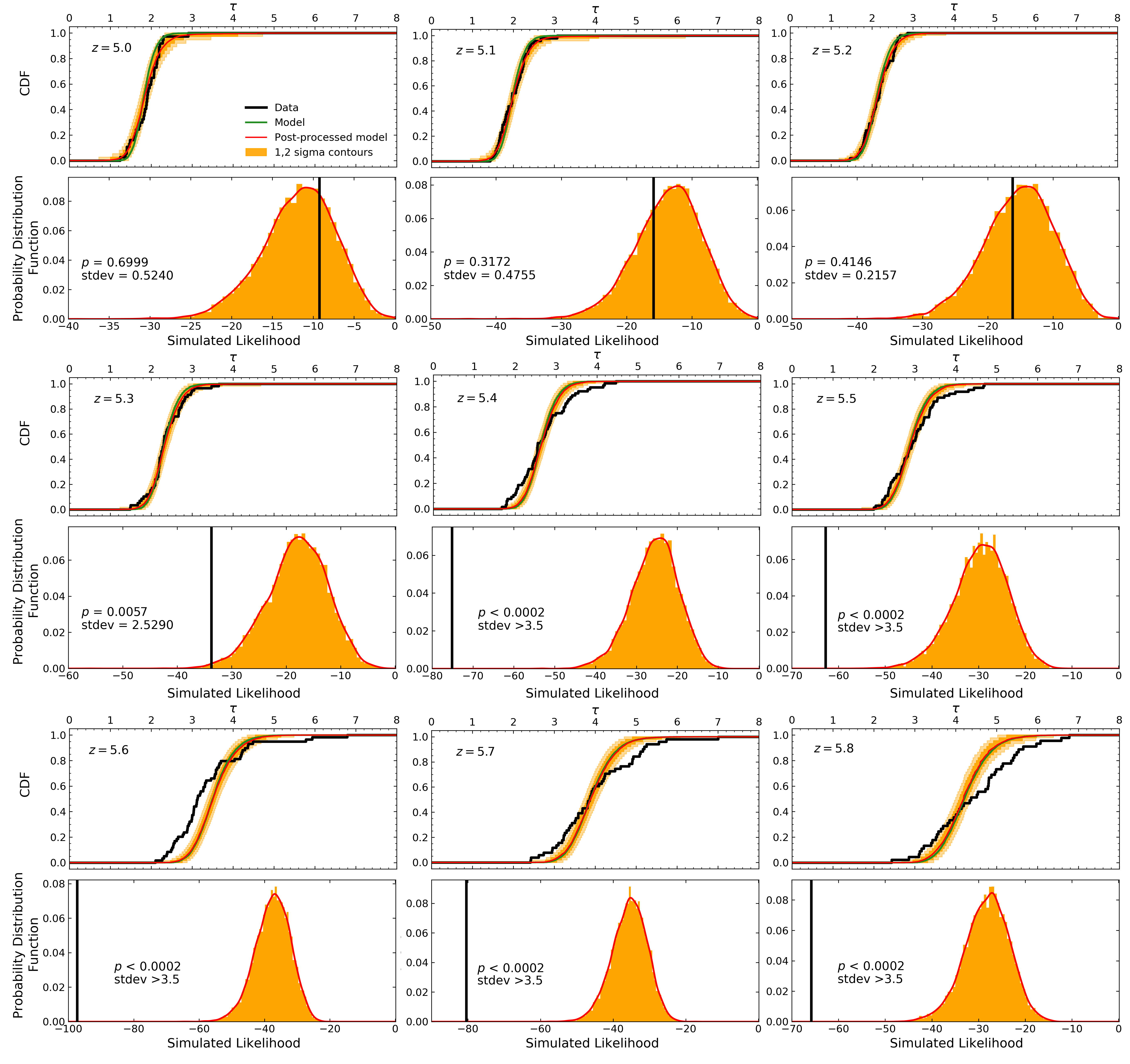}
\caption{Top panels: observed \lal optical depth distribution (black), compared to the Sherwood simulations without any post-processing (green) and with post-processing (red, orange). The light and dark orange contours show $1\sigma$ and $2\sigma$ envelopes from bootstrap resampling the post-processed models. Bottom panels: probability distribution of log-likelihoods for fully forward-modelled datasets (orange). The distributions are used to build KDEs (red) which are evaluated at the location of the likelihood of the observed dataset (thick vertical black lines). The Sherwood model is a great fit to observations at $z\leq5.2$, but excluded at $>3.5\sigma$ at $z\geq5.4$.}  \label{fig:Sherwood}
\end{figure*}

\begin{figure*}
\includegraphics[width=\textwidth]{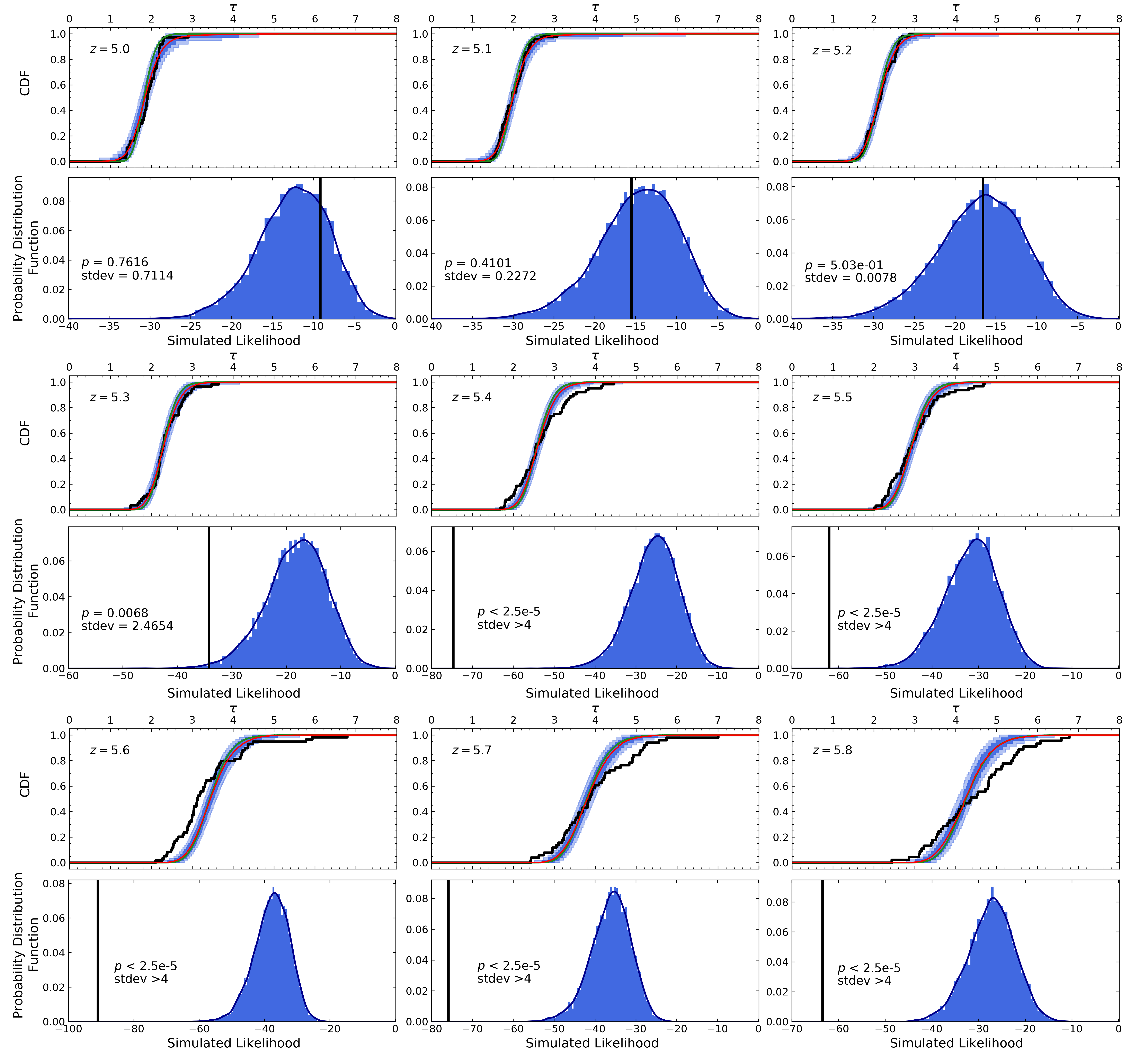}
\caption{Same as Figure~\ref{fig:Sherwood} but for the Nyx simulation (blue). The Nyx model is a great fit to observations at $z\leq5.2$, is in mild tension with data at $z=5.3$ ($\revmath{2.5}\sigma$) and excluded at $<3.5\sigma$ at $z\geq5.4$. }  \label{fig:Nyx}
\end{figure*}

\section{Comparison with homogeneous-UVB models}\label{sec:models}

The existence of completely opaque troughs with $\tau>5$ at $z\sim5.8$ rules out models of reionisation with a homogeneous UVB and IGM temperature-density relation \citep{Becker15, Bosman18}. 
Even in the absence of large opaque troughs, the observed scatter in \lal optical depth at fixed redshift suggests excess fluctuations at even later times (Fig.~\ref{fig:large}). 
Determining the redshift evolution of these fluctuations can quantify the transition redshift beyond which the IGM no longer retains reionisation-related structure from the point of view of \lal absorption.

We compare our results to predictions from two different homogenous-UVB simulations, Sherwood \citep{Bolton17} and Nyx \citep{Nyx}. In both models, scatter between sightlines results solely from fluctuations in the density field within a constant, fully-permeated UVB. Compared to Nyx, the Sherwood simulation is run with smaller boxes but provides finer redshift sampling ever $\Delta z=0.1$ in redshift, while in Nyx the optical depth distribution must be extrapolated from $3$ snapshots at $z=5.0$, $z=5.5$ and $z=6.0$. The two simulation suites also employ different models of the UVB with different base rescalings of the ionising intensity. Neither simulation resolves the gas densities corresponding to DLAs nor Lyman-limit systems. We give more details of the simulation suites below.

The Sherwood simulation suite was designed to reproduce \lal transmission post-reionisation, at $2<z<5$, where it is in remarkable agreement with observations \citep{Bolton17}. Sherwood employs the hydrodynamics code P-GADGET 3 \citep{Springel05} and a uniform \citet{HM12} UVB. The gas particle masses are $M_\text{gas} = 9.97 \cdot 10^4 M_\odot$ and the box includes  $2\times 2048^3$ particles. 
We use the simulated boxes which are $40$ cMpc/h on the side; we prefer those boxes over the lower-resolution $80$ cMpc/h runs of Sherwood since they resolve the \lal transmission and thus provide a closer comparison to the Nyx simulation. 
Snapshots were taken every $\Delta z = 0.1$ from $z=5.0$ to $z=6.0$. We draw $5000$ lines of sight through the simulation box with lengths corresponding to $\Delta z=0.1$ at each redshift. 

Nyx is an Eulerian grid cosmological hydrodynamical simulation code which is optimized for simulations of the Ly$\alpha$ forest \citep{Lukic15}. We use the Nyx simulation described in \citet{Davies18-spikes}, $100$ cMpc/h on a side with $4096^3$ dark matter particles and $4096^3$ baryon grid cells, sufficient box size and resolution for converged Ly$\alpha$ forest statistics at $z\lesssim6$ \citep{Onorbe17}. Snapshots at $z=5.0$, $5.5$, and $6.0$ were used to simulate the Ly$\alpha$ forest at $z=5.0$--$5.2$, $5.3$--$5.7$, and $5.8$, respectively. At redshifts not equal to the snapshot redshift, we re-scaled the physical gas densities by $(1+z)^3$ to account for cosmological expansion, effectively ignoring the impact of structure formation over these intervals of cosmic time. We draw $40000$ lines of sight through each snapshot starting from random positions within the volume towards a random direction along the grid axes. While the simulation was originally run with the \citet{HM12} UVB for heating and cooling rates, here we initially construct Ly$\alpha$ forest skewers assuming a fixed \revi{photo-ionisation} rate $\Gamma_{\rm HI}=10^{-12.1}$\,s$^{-1}$ comparable to observational estimates at $z\sim5$ \citep{Becker13-UVB}.


We post-process sightlines drawn from simulations in the following way. First, we shorten sightlines to the length corresponding to $\Delta z = 0.1$ and project them onto a wavelength array with constant velocity sampling. We then randomly assign each simulated sightline to a real observation in the same redshift interval, and interpolate the simulated flux onto the observed wavelength array (including any masking of bad regions). We add random noise sampled from the corresponding observed error array by drawing from a gaussian with width of the $1\sigma$ uncertainty at each pixel. Finally, we multiply the sightline by a wavelength-dependent continuum error drawn from a normal distribution with scale of the observed $1\sigma$ bound of the continuum uncertainty. This assumes that the continuum uncertainty is fully covariant, while formally we would need to draw from the full PCA posterior (see e.g.~\citealt{Davies18-DW}). However, since we care not about the details of wavelength-dependence, our approach is both more computationally efficient and conservative. Shifting the continuum reconstruction at all wavelengths by the same standard deviation introduces a more coherent shift than selecting a random draw with the same PCA likelihood. This procedure will tend to introduce ``pessimistic'' continuum-reconstruction scatter into the post-processed simulations, in the sense that it slightly lower the evidence for fluctuations (which is conservative for our purposes). 
Note that we do not need to convolve our simulated sightlines to match the observed instrumental resolutions, since the convolution operation explicitly conserves the total flux.




\subsection{Maximum likelihood analysis}\label{sec:ml}

In order to calculate the likelihood of our observations given a model, optical depths at each simulated pixel must first be rescaled in bulk. The optical depth rescaling is expressed as a multiplicative factor on optical depth, $\tau_\text{rescaled} = A \tau_\text{sim}$ where the rescaling factor $A$ is different at each redshift\footnote{Note that this rescaling is applied at each simulated pixel before computing the mean optical depth.}. Rescaling corresponds to adjusting the ionising background intensity in the simulations, i.e.~\revi{for Sherwood} it reflects a deviation from \citet{HM12} in the average ionising emissivity which can be a factor of a few. Optical depth rescaling factors are usually chosen to match the observed mean fluxes at each redshift, $\left< e^{A \tau_\text{sim}} \right> = \left< F \right>_\text{obs}$, but this may lead to bias when large sightline scatter is present. A few highly transmissive sightlines will lead to a very low average $\tau$, which might make it difficult to match opaque sightlines. However, the highly transmissive sightlines also carry uncertainties. Therefore, a rescaling to a slightly lower average flux than observed leads to a better agreement between models and observations, because both opaque and transmissive sightlines can be produced by random noise. Motivated by this observation, we choose the rescaling factor to \textit{maximise the likelihood of the observations} instead of matching the mean flux explicitly. We note our rescaling factors still give rise to mean simulated fluxes consistent with observed mean fluxes at $<1\sigma$ at all redshifts where the simulations are a good fit to the data (see below).

We determine the likelihood of the observations by combining the likelihoods of each individual measurement $\tau_n$ made in sightline $S_n$:
$$
\mathcal{L}_\text{data} = \prod_{n=1..N} p(\tau_n | S_n)
$$
where $N$ is the total number of sightlines contributing to the distribution at a given redshift. The probability $p(\tau_n | S_n)$ is obtained by post-processing all simulated sightlines with the observational properties of sightline $S_n$: wavelength sampling and masking, random flux uncertainties, and a random continuum uncertainty. The resulting distribution of predicted $\tau$ given $S_n$ is then used to build a kernel density estimator (KDE). To obtain a smooth KDE from the Sherwood simulation with a relatively small number of sightlines, we over-sample each sightline $6$ times\footnote{meaning the observational uncertainties are chosen at random $6$ times for each sightline to produce $6$ predicted values of $\tau$.}. The KDE is then evaluated at the observed value $\tau_n$ to produce $p(\tau_n | S_n)$. The process is repeated for each observation $S_n$ to obtain $\mathcal{L}_\text{data}$ via Equation (2). 

For the purposes of comparison with models, we always assume that flux non-detections correspond to intrinsic flux just below the detection threshold (i.e.~the upper CDF bounds in Figure~\ref{fig:dz}). 
\revi{We thus ensure that simulations are given the `best possible chance' at reproducing the optical depth scatter in the observations, since homogeneous-UVB models are known to always under-estimate (and never over-estimate) \lal optical depth scatter. Non-detections of mean flux only occur at $\revmath{z\geq5.6}$, therefore this definition is equivalent to using the measured values of flux at all redshifts where the models are a good description of the observations.} 
We pick the rescaling factor to maximise $\mathcal{L}_\text{data}$ by sampling $A$ in steps of $0.0025$. 

\revi{Table~\ref{table:resc} gives the optimally-rescaled photo-ionisation rates, $\Gamma = \Gamma_\text{sim}/A$. The Sherwood simulation required rescaling $\Gamma$ down by up a factor of $2$ at $z\sim5.0$ ($A=0.5$) while Nyx's photo-ionisation rate was rescaled up by up to a factor $2$ at $z\sim6.0$ ($A=2$). After rescaling, $\Gamma$ is in good agreement between the simulations, within $\sim12$\%. This remaining difference at $z\leq5.3$ can most likely be attributed to small differences in cosmological parameters (e.g.~$\Omega_m$, $\sigma_8$) between the simulations, to which $\Gamma$ is known to be sensitive \citep{Bolton07}. $\Gamma$ shows a discontinuity in Nyx at $z=5.3$ due to switching from extrapolating the $z=5.0$ snapshot to the $z=5.5$ snapshot; the difference between using the two snapshots is about $20\%$. Any tensions are far below current measurement uncertainties in $\Gamma$ due to the IGM's thermal state which are a factor of $\sim 2$ (e.g.~\citealt{Daloisio18}).}

We now calculate the probability of drawing a full dataset with $\mathcal{L}_\text{data}$ from the simulations. We generate $10000$ fully forward-modelled datasets by post-processing $N$ randomly-selected model sightlines, each assigned to the uncertainties of an observed sightline $S_n$. All simulated datasets therefore have the same size as the observations. The likelihood is calculated for each simulated dataset in the same manner as the data\footnote{However, we do not apply the optimal choice of rescaling factor to each simulated dataset as for the data, since this would be computationally unfeasible. A test of the impact reveals that the wings of the likelihood distribution may shift to higher values by up to $\Delta\mathcal{L}\sim2$, which is insufficient to quantitatively affect our results.}, giving rise to a distribution of $\{ \mathcal{L}_\text{sim} \}$. We build a KDE on the distribution of simulated likelihoods and evaluate it at $\mathcal{L}_\text{data}$ to finally obtain the probability of the entire set of observations given the simulation model. These probabilities $p$ formally coincide with the p-values, and we also convert them to standard deviations via $\text{stdev} = \sqrt{2} \text{erf}^{-1}(2 p)$ where $\text{erf}^{-1}$ is the inverse error function.


\subsection{Results}

Figures~\ref{fig:Sherwood} and \ref{fig:Nyx} show the results of the likelihood analysis for the Sherwood and Nyx simulations, respectively. The data likelihood falls within $\pm 1\sigma$ expectations at $5.0\leq z \leq 5.2$ for both Sherwood and Nyx. Forward-modelling introduces some optical depth scatter due to uncertainties, most visible at $z=5.0$. The extra scatter is expected, and provides a better fit to observations: e.g.~at $z=5.1$ and $z=5.2$ in Sherwood, the post-processed elongated distribution (red line) provides a better fit to the data than the model without post-processing (green line). The excellent agreement with models at $z\leq5.2$ implies that the intrinsic physics within the simulations combined with our known observational uncertainties account for all the variance observed in the data. A homogeneous UVB acting on density fluctuations is therefore a sufficient description of \lal transmission up to $z=5.2$.

Conversely, the \lal transmission scatter observed at $z\geq5.4$ is in excess of model predictions at $>3.5\sigma$ in both models. Since we sampled $5000$ sightlines from the Sherwood simulation, we are limited in determining the nature of outliers to the $\lesssim 1/5000 \simeq 3.5\sigma$ level. The $40000$ sightlines from the Nyx simulation enable us to push the analysis to $\lesssim 1/40000 \simeq 4\sigma$ outliers. We find that the post-processed Sherwood simulations fail to match the observations at the saturation level ($3.5\sigma$) at all redshifts $z\geq5.4$. Nyx similarly fails to match observations at the corresponding $4\sigma$ level at $z\geq5.4$. 
In both models, the rescaling factor which maximises the likelihood of observations results in mean simulation fluxes in close agreement with observed values (within $1\sigma$ of the values in Table~\ref{table:means}) at $z\leq5.3$; but the mean fluxes are in disagreement at $z\geq5.4$ where the `most likely' mean fluxes are closer to the median (Figs~\ref{fig:Sherwood}, \ref{fig:Nyx}). As expected, matching the median transmission increases the likelihood of a extended distribution since both extremely opaque and extremely transmissive sightlines then have reasonable probabilities.

\begin{figure}
\includegraphics[width=\columnwidth]{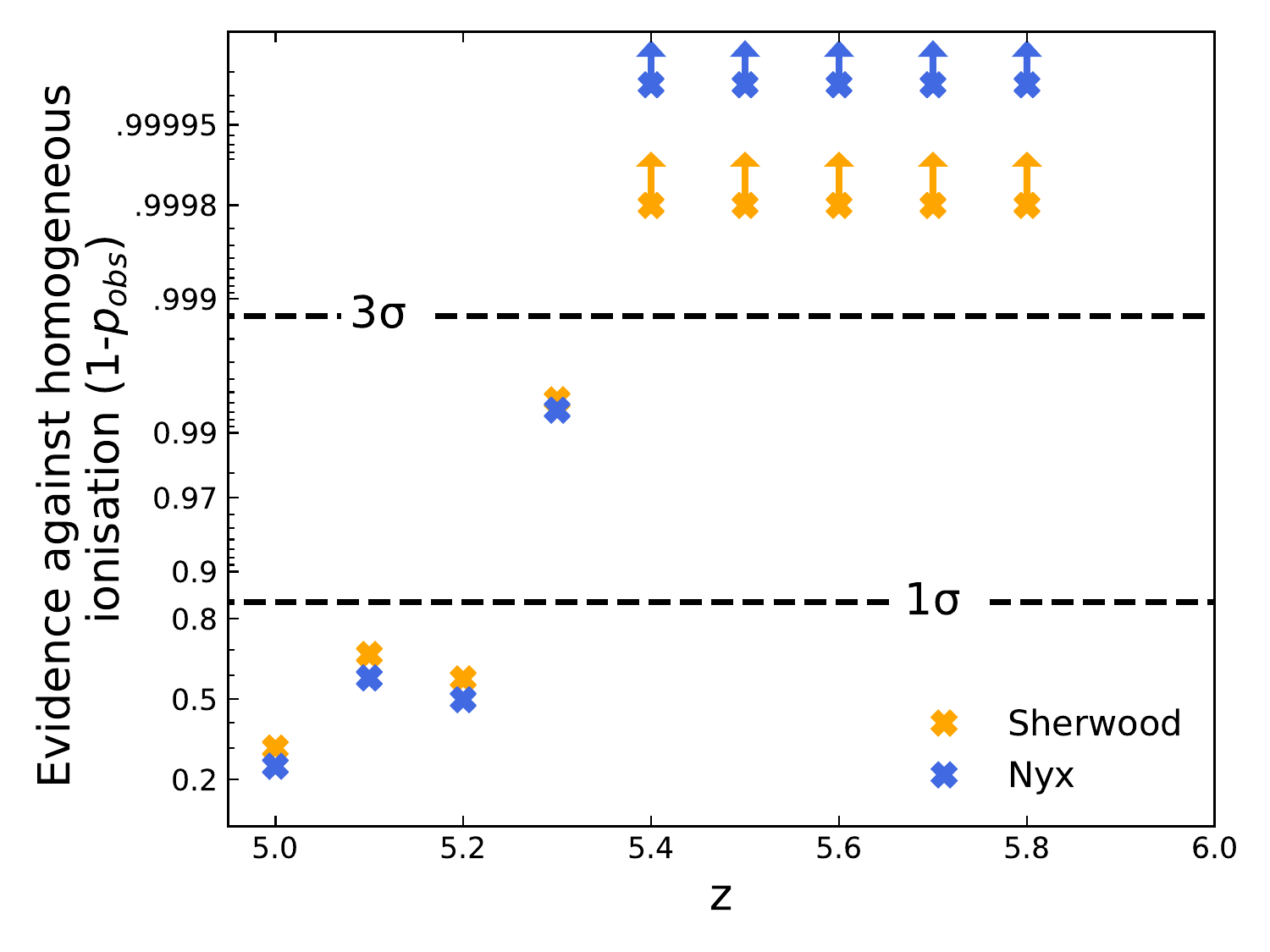}
\caption{Summary of the likelihood analysis comparing the observed distributions of optical depth to the homogeneous-UVB Sherwood (orange) and Nyx (blue) simulations. By including forward-modelling of all known uncertainties, both models provide an excellent fit to the data at $z\leq5.2$ but are in strong tension at $z\geq5.4$.}  \label{fig:tension}
\end{figure}





Both the Nyx and Sherwood simulations transition from providing good fits to the data to being in strong tension with observations at $z=5.3$, where they are in mild tension with the data (excluded at $2.4\sigma$ and $2.5\sigma$, respectively). 
We conclude that extra \lal optical depth scatter is present in the observations, and its magnitude is in excess of differences between simulations due to box size and different choices of UVB models. However, the tension is mild and we cannot completely rule out either homogeneous UVB model. Examining the difference between Nyx and Sherwood in more detail, we find that Nyx provides a statistically better fit to the data at all redshifts in the absolute (i.e.~the likelihood of the observed dataset is higher in Nyx). We attribute this to Nyx's larger box size, which makes the model more apt to capture density fluctuations on large scales. However, both models are in agreement with the data at $z<5.3$, in strong tension at $z>5.3$, and in mild tension at $z=5.3$.




Redshift $z=5.4$ is the lowest redshift at which the observed distribution of optical depths is in strong tension with both models ($>3.5/4\sigma$). To evaluate the robustness of the result, we test whether the tension is driven primarily by the most opaque sightlines at $z=5.4$ or by the extended shape of the entire distribution. We arbitrarily remove the most opaque $3$ sightlines, which have observed $\tau>4$ -- corresponding to $5\%$ of the sample. While none of them show signs of foreground absorption by DLAs in the form of intervening metal absorbers, some DLAs at $z>5$ may be particularly metal-poor. Even though we find no evidence for such metal-poor DLAs in the other redshift bins, unlucky alignment cannot be completely excluded. We roughly estimate that such DLAs would need metallicities of $\left[ \text{X}/\text{H} \right]\lesssim -2.5$ to avoid detection in our spectra; this will be calculated in more detail in \revi{future work}.

However, we find that even after arbitrarily removing the most opaque $3$ sightlines from the distribution, the observations are still in strong tension with the $z=5.4$ Sherwood simulation at $3.43\sigma$ ($p=0.0305\%$ with $A=0.58$). A similar result is obtained with Nyx, where omitting the most opaque sightlines still results in tension at $3.68\sigma$. We show the corresponding likelihood distributions in Appendix~\ref{app:wo}. We therefore conclude that the whole extended shape of the distribution, and not just a few sightlines, are driving the disagreement between homogeneous UVB models and observations at $z=5.4$.

\subsection{Discussion}

Figure~\ref{fig:tension} summarises the results of the likelihood analysis. Both homogeneous-UVB simulations, Sherwood and Nyx, provide an excellent fit to observations at $z\leq5.2$. Post-processing the simulations slightly broadens the predicted distribution of optical depths in this regime, bringing predictions in agreement with the data. There is no evidence that any extra sources of fluctuations are necessary at $z\leq5.2$, such as, for example, a spatially varying thermal state of the IGM. In particular, the Sherwood simulation successfully matches \lal optical depth over $2\leq z\leq5.2$ without any such modifications \citep{Bolton17}. The tension observed at $z\geq5.4$ is therefore highly significant, and marks the breakdown of one or more simplifying assumptions in the post-reionisation high-$z$ IGM.


\revi{A potential caveat to our maximum likelihood analysis is that the statistical power of the homogeneous-UVB simulations may be limited by box size rather than by the number of simulated sightlines.} 
\revi{Indeed, the Nyx simulation box only contains $\revmath{\sim 50}$ independent volumes of scale comparable to the lengths of observed sightlines ($\revmath{\sim 30-35}$ cMpc/h) while the Sherwood box contains only a few. This is much lower than the $\revmath{10000}$ independent draws necessary to establish statistical significance at the $\revmath{4\sigma}$ level. However, the fact that the two simulations result in very similar large-scale optical depth CDFs suggests that the modes of the density field which dominate the large-scale opacity fluctuations are actually much smaller than the total path length which should thus be much better sampled (see e.~g.~the Appendix of \citealt{Becker15}, who found that $\revmath{50}$ Mpc/h-scale fluctuations were extremely similar between $\revmath{50}$ Mpc/h and $\revmath{100}$ Mpc/h simulation volumes). 
Treating the sampling as being limited by the number of independent large-scale modes would therefore be somewhat too conservative. Nevertheless, our analysis is only strictly valid in the context of the specific simulation boxes we used.} 
\revi{While our results suggest that density fluctuations on scales larger than $\revmath{\gtrsim 40}$ cMpc/h play a negligible role in determining the \lal optical depth at $\revi{z<5.3}$, we note that larger simulated volumes are crucial to modelling the reionisation process at higher redshifts, especially in models where bright rare sources play a significant role (e.g.~\citealt{Chardin17, Meiksin20}).}

The presence of large opaque troughs $\gtrsim 100$ cMpc/h in length in the \lal forest down to $z\sim5.6$ already independently rules out homogeneous ionisation at that redshift \revi{\citep{Becker15, Bosman18, Zhu21}}. Opaque troughs persisting at late times have been theorised to arise from patches of significantly neutral gas ($x_\text{HI}>10\%$, \citealt{Kulkarni19, Keating20, Nasir20}; see also \citealt{Lidz06,Mesinger10}). At the same time, recent measurements have reported a very short mean free path of ionising photons at $z=6$, of $\sim0.75$ pMpc \citep{Becker21}. Evidence therefore points to a late end of reionisation, with remnant fluctuations in the UVB and/or IGM temperature persisting for \revi{at least} $70-80$ Myr after the demise of the last highly neutral `patches' at $z\simeq5.6$ (see also \citealt{Davies21, Cain21}). From the point of view of \lal transmission homogeneity, hydrogen reionisation is not over before $z=5.3$.

\begin{figure}
\includegraphics[width=\columnwidth]{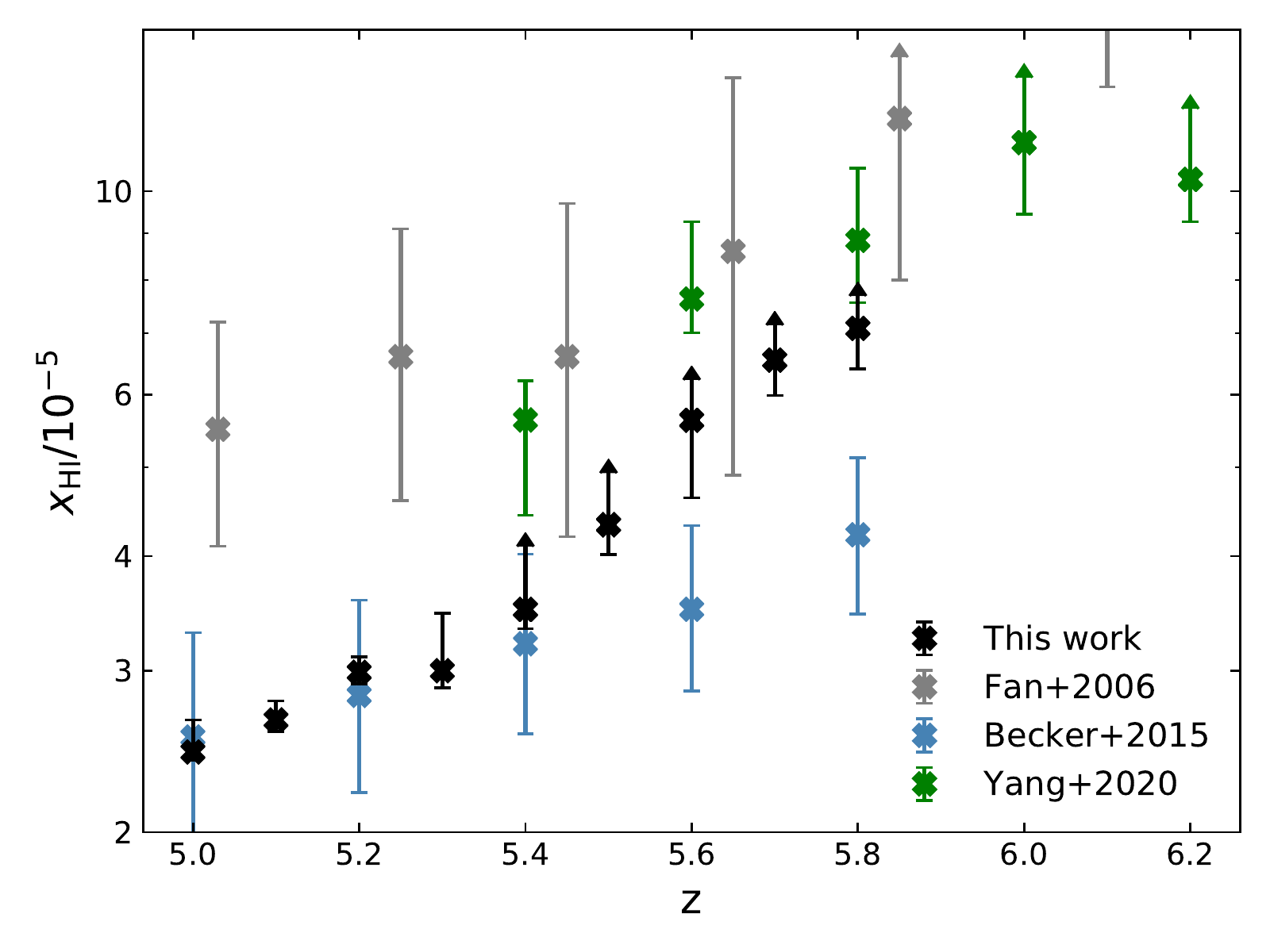}
\caption{Evolution of the volume-averaged neutral fraction $x_\text{HI}$ with redshift. We only provide lower limits in the regime where the Nyx simulation, which we use in the calculation of $x_\text{HI}$, is excluded by the observations. The model is a fairly poor fit to the data at $z=5.3$ ($2.5\sigma$ tension), which may be causing an offset. Our measurements employ samples factors $5-10$ larger than past measurements. Tension with the \citet{Fan06} values (blue) may be the result of a number of observational, systematic, or modelling differences.}  \label{fig:xhi}
\end{figure}

\subsubsection{Hydrogen neutral fraction}


We calculate the volume-averaged $x_\text{HI}$ directly from the $40000$ Nyx skewers at each redshift assuming ionisation equilibrium, shown in Figure~\ref{fig:xhi}. 
Our nominal measurements correspond to re-scalings of the UVB which maximize the likelihood of the $\tau_\text{eff}$ distribution (Table~\ref{table:xhi}). We also measure upper and lower bounds corresponding to re-scalings that reproduce, respectively, the lower and upper bounds of the mean transmitted flux. 
At $z=5.3$, we find a significant difference between the $x_\text{HI}$ values obtained from rescaling the $z=5.0$ snapshot of Nyx as opposed to the $z=5.5$ snapshot; we attribute this difference to evolution in the cosmic structure which neither snapshot captures perfectly. We list the most pessimistic bounds on $x_\text{HI}$ among both snapshot re-scalings. 
For redshifts $z\geq5.4$, the strong tension between our data and the maximum-likelihood Nyx $\tau_\text{eff}$ distribution implies that reionisation may not yet be complete (see also \citealt{Kulkarni19,Nasir20,Qin21, Choudhury21}). The $x_\text{HI}$ estimate from the mean flux is not sensitive to the fraction of fully-neutral regions, so we show our measurements at $z\geq5.4$ as lower limits. \revi{Since the homogeneous-UVB simulations are rescaled to maximise the likelihood of `optimistic' observations where non-detections are treated as flux just below the detection limit (Section \ref{sec:ml}), our $\revmath{x_\text{HI}}$ limits might be too conservative by $\revmath{\lesssim 5\%}$. This effect goes in the same direction as the lack of inclusion of self-shielding in the models. }

The calculation of $x_\text{HI}$ has traditionally assumed an optically-thin IGM without self-shielding by dense fluctuations \citet{Fan06, Becker15, Yang20}. To estimate the impact of this assumption on our measurements, we post-process a set of skewers with the prescription of \citet{Rahmati13} and show the results in Table~\ref{table:xhi}. The inclusion of self-shielding results in an increase of $x_\text{HI}$ by $\sim25\%$. Unlike previous works where the effect was comparatively negligible, uncertainties related to the treatment of self-shielding dominate over our statistical uncertainties. Since the Nyx simulations do not resolve dense gas, we cannot provide a physically realistic inclusion of self-shielding at the $<5\%$ level required to match the statistical uncertainties. We show the best-fit values without self-shielding in Figure \ref{fig:xhi} in order to compare to past work which universally assumed an optically-thin IGM.

We are consistent with the inferred $x_\text{HI}$ values of \citet{Yang20}, who employed a homogeneous UVB model up to $z=5.8$. Our values of the neutral fraction at $5.0\leq z\leq5.4$ are a factor $\sim2$ lower than reported by \citet{Fan06}. This tension ($\sim2\sigma$) may be due to a number of factors, such as a much smaller sample size than our study, continuum reconstruction systematics, lower SNR, or the very significant differences in the IGM model. 
We are in good agreement with \citet{Becker15} up to $z=5.4$. The measurements of \citet{Becker15} correspond to the neutral fraction specifically \textit{inside of ionised regions}, which explains the divergence with our lower limits at higher redshifts. 
The uncertainties of our low-$z$ $x_\text{HI}$ measurements are very small, reflecting the exquisite precision of the measurement of the mean flux (Fig.~\ref{fig:means}). We warn that homogeneous UVB models are a fairly poor fit to observations at $z=5.3$ ($2.5\sigma$ tension), such that systematic errors in $x_\text{HI}$ may be present in that redshift bin. The use of a $-1\sigma$ lower bound on $x_\text{HI}$ as a lower limit may therefore be an equally justified choice.

The conversion of mean flux measurements to values of the IGM neutral fraction $x_\text{HI}$ is only valid under the assumption of completely homogeneous ionisation. This is because, fundamentally, the translation is model-dependent and relies on simulations assuming homogeneous ionisation. Models which reproduce the mean flux with late reionisation, such as that of \citet{Kulkarni19}, predictably result in significantly higher $x_\text{HI}$ than simulations with homogenous ionisation even when they match \textit{the same observed mean flux} \citep{Yang20}.

\begin{table}
\centering
\begin{tabular}{l c c}
$z$ & $x_\text{HI}/10^{-5}$ (no s-s) & $x_\text{HI}/10^{-5}$ (with s-s) \\
\hline
$5.0$ & $2.446 -0.051   +0.205$ & $3.020  -0.058 +0.230$\\
$5.1$ &  $2.651 -0.075 +0.129 $ & $3.336  -0.164 +0.064$\\
$5.2$ &  $2.988 -0.085 +0.119 $ & $3.636  -0.095 +0.131 $\\
$5.3$ &  $3.000 -0.125 +0.466$ & $ 3.598  -0.145 +0.566$ \\
$5.4$ & $(3.498) \ >3.332$ &$-$\\
$5.5$ & $(4.328) \ >4.016$ &$-$\\
$5.6$ &$(5.627) \ >4.630$ &$-$\\
$5.7$ & $(6.544) \ >5.990$ &$-$\\
$5.8$ & $(7.087) \ >6.401$ &$-$\\
\hline
\end{tabular}
\caption{\revi{Volume-averaged hydrogen neutral fraction} and $\pm1\sigma$ bounds computed by comparison with the homogeneous-UVB Nyx simulations. We show the values both with and without self-shielding included (s-s). The numbers in brackets have the highest likelihood according our model, but note that above $z\geq5.4$, homogeneous-UVB simulations are a poor match to data and only enable lower limits on $x_\text{HI}$. }\label{table:xhi}
\end{table}

\begin{figure*}
\includegraphics[width=\textwidth]{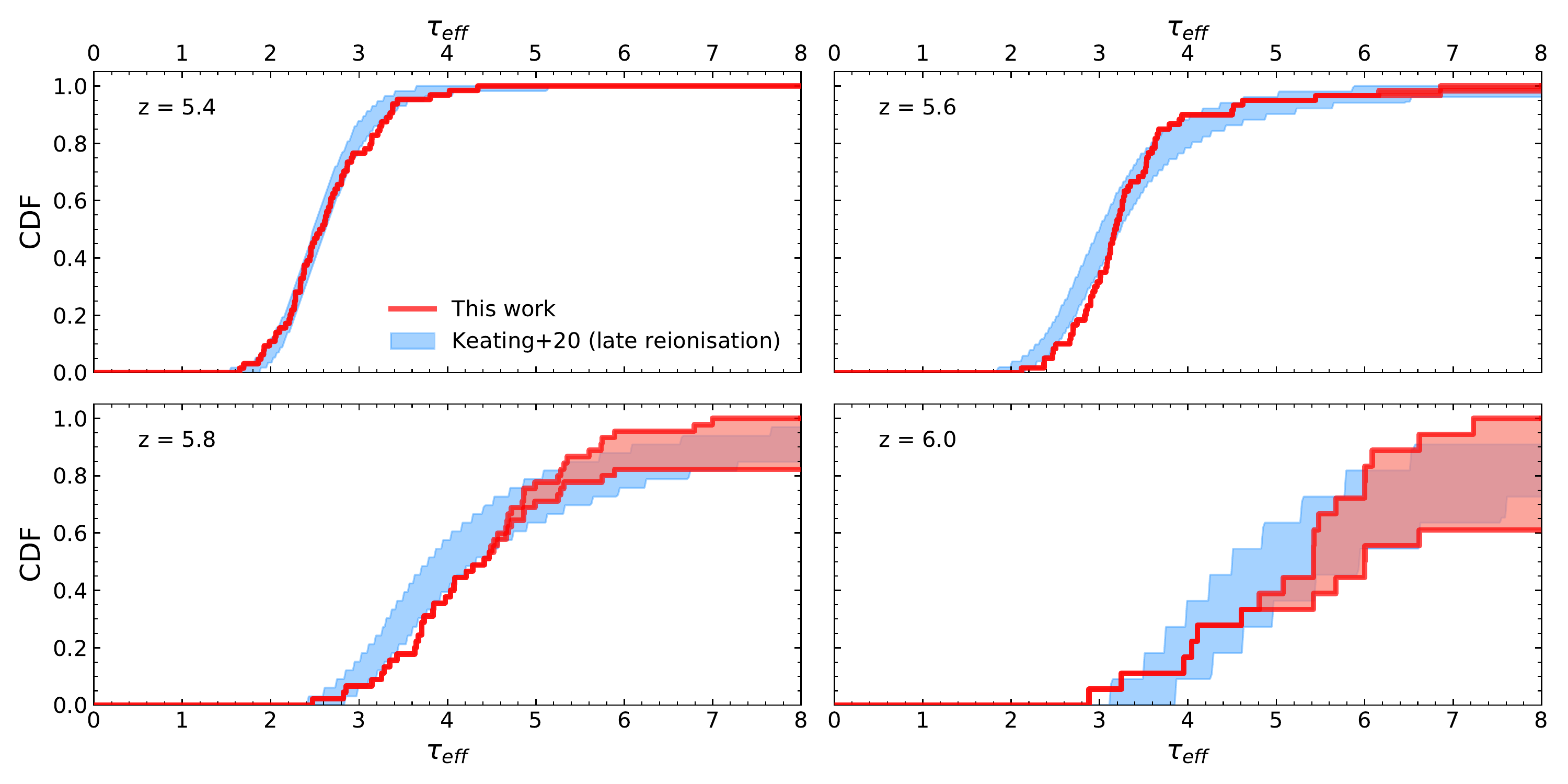}
\caption{Comparison of the optical depth distributions measured in this work (red) to a model of late reionisation (\citealt{Keating20}, blue). The contours of the blue distribution correspond to the central $1\sigma (70\%)$ bounds of the simulated distributions. Late reionisation provides an excellent qualitative description of the \lal optical depth scatter above $=5.4$. The late reionisation model was not calibrated to reproduce our updated measurements to mean \lal transmitted flux, precluding a direct quantitative comparison.}  \label{fig:Keating}
\end{figure*}

\subsubsection{Non-homogeneous UVB models}

We compare our new measurements of \lal optical depth distributions with the late-reionisation model of \citet{Keating20} (first described in \citealt{Kulkarni19}). Their model uses high-resolution cosmological radiative transfer simulations in boxes of $160$ cMpc/h on the side. Similarly to the Sherwood suite, the late-reionisation simulation is run with the P-GADGET 3 code and uses the same cosmological initial conditions. The radiative transfer is conducted in post-processing with the ATON code \citep{Aubert08, Aubert10}. The simulation employed $2\times 2048^3$ gas and dark matter particles. Lightcones of $50$ cMpc/h are extracted from the simulation on-the-fly, resulting in sightlines with H~{\small{I}} fraction and temperature that evolve along the line of sight with redshift. The centre of each such sightline is matched to the mid-point redshift of a measurement from the \citet{Bosman18} sample, such that all sightlines are at slightly different redshifts. Five hundred such simulated datasets are generated for each redshift. Figure~\ref{fig:Keating} shows the resulting $1\sigma$ ($70\%$) bounds of the corresponding CDFs. 

In order to compare these simulations to our observations, we re-bin the spectra in intervals of $50$ cMpc/h centred on the mid-point redshift of each snapshots. The resulting bins cover redshifts of $5.3245<z<5.4754$ for the $z=5.4$ snapshot, $5.5210<z<5.6710$ for the $z=5.6$ snapshot, etc. In addition to the like-to-like sightline matching detailed above, the predicted \lal optical depth distributions from \citet{Keating20} were also calibrated to the mean \lal transmitted flux measurements of \citet{Bosman18}. The late reionisation model cannot be trivially re-scaled to our updated mean flux values, because the radiative transfer simulations rescale the emissivity of reionising sources to match the mean flux and predict \lal optical depth fluctuations self-consistently. As such, the model requires time-consuming runs of the simulation to calibrate. We therefore leave a quantitative comparison of the late reionisation model with our observations to future work. Figure \ref{fig:Keating} shows the excellent qualitative agreement between our new observations and the \citet{Keating20} model without rescaling to match the new measurements of mean transmitted flux, nor sightline matching. At $z=6.0$, the late reionisation model predicted a significantly larger fraction of sightlines $\tau_\text{eff}< 4$ than observed in previous work (compare with Figure~\ref{fig:compar}). At $z=5.8$, the model also predicted a larger number of highly opaque sightlines, closer to our present measurements than to previous measurements. The agreement with our updated results is therefore excellent despite the lack of specific re-calibration. This is encouraging evidence for a patchy, late end to hydrogen reionisation.



\section{Conclusions}\label{sec:ccl}

We have measured the mean \lal optical depth at $4.8<z<6.2$ by assembling a sample of $67$ high-SNR quasar sightlines, leveraging the new XQR-30 sample of X-Shooter spectra of $z\gtrsim5.8$ quasars. Our sample represents a $\sim3$-fold increase in the number of high-quality spectra of \lal transmission at the end stages of reionisation. 
We only employ observations taken with $2$ spectrographs, enabling us to rigorously quantify systematics in instrumentation and continuum reconstruction for all our observations. The depth of observations, SNR $>10$ per spectral pixel, also enables a more careful removal of possible DLA contaminants than previous studies. 

Our measurement of the evolution of the mean \lal evolution with redshift is in rough agreement with previous work (Figure~\ref{fig:means}). Differences are more likely to originate in previously-uncorrected systematics than in cosmic variance, given our large sample size and overlap with previous studies. We detect no sudden acceleration in the mean flux evolution over $4.8<z<5.5$. 

We present an extremely transparent sightline with $\tau<3$ at $z=5.9$, and $2$ rare sightlines with $\tau<4$ at $z=6.1$. These rare sightlines correspond to patches of the IGM with factors $5-15$ times more transmitted flux than the median. The existence of transparent patches may help constrain future models of reionisation, which must be able to generate both sightlines with $\tau\sim2.5$ and $\tau>6$ at the same redshift ($z=5.9$).

Next, we determine the lowest redshift at which excess optical depth scatter in \lal emerges, signalling a departure from a uniformly ionised IGM. Using an improved grasp on systematics, we forward-model two simulation models employing homogeneous UVBs, the Sherwood and Nyx simulations. We conduct a maximum-likelihood analysis to obtain the probability of the full observed dataset at each step of $\Delta z = 0.1$. All observational systematics (wavelength masking, observational uncertainties, continuum uncertainties, etc) are included in post-processing of the simulations. These uncertainties result in increased \lal optical depth scatter which improves the agreement between models and observations. 

We find excellent agreement between the forward-modelled simulations and observations at $5.0\leq z\leq5.2$, where the observed data has a high probability of being observed by chance ($<1\sigma$). A homogeneous UVB is in mild tension with observations at $z=5.3$ ($2.5\sigma$) and strongly excluded at $z\geq5.4$ ($>4\sigma$). To check whether the disagreement at $z=5.4$ is driven by a few opaque sightlines which may contain DLAs, we arbitrarily remove the $3$ least transmissive sightlines which have $\tau>4$. Homogeneous UVB models remain excluded at $\revmath{z=5.4}$ at $>3.5\sigma$ confidence, meaning that the intrinsically large width of the observed distribution, and not just a few sightlines, is driving the tension. Despite differences in the box size, snapshot density, and UVB models between the two suites, our results are consistent between the Sherwood and Nyx simulations. 

Since the Sherwood model has been highly successful in modelling the \lal forest over a wide range of redshifts ($2<z<5.2$), a sudden failure by $z=5.4$ represents a breakdown of one or more simplifying assumptions. Whether fluctuations in the UVB are present at very late times and/or whether the thermal state of the IGM retains the imprint of recent ionisation, it is clear that reionisation-related fluctuations persist in the IGM until at least $z=5.3$.

Finally, we convert our measurements of the mean \lal flux to volume-averaged neutral fraction $x_\text{HI}$. We stress that this conversion is model-dependent; here we use the Nyx simulation suite. Since Nyx (and homogeneous-UVB models in general) provides a very poor fit to data at $z\geq5.4$, only lower limits on $x_\text{HI}$ can be quoted. Our results at $5.0\leq z \leq 5.3$ are in mild tension with those or \citet{Fan06} ($\sim2\sigma$), but the vast improvements in data quality, quantity and understanding of systematics and IGM modelling over the last $15$ years makes it difficult to pinpoint the source of the disagreement. 

The XQR-30 sample has qualitatively changed the landscape of the late stages of reionisation. Analysis of \lal transmission at $z>5$ has become a precision probe of the post-reionisation era, with exciting prospects both on the analysis and theoretical fronts. Through excellent complementarity with upcoming 21cm probes, IGM transmission studies make it possible to uncover the entire history of reionisation from start to end.


\section*{Data Availability}

The quasar spectra used in this analysis will be shared on reasonable request to the corresponding author. The XQR-30 spectra and associated meta-data will further be made public in the upcoming data release of D'Odorico et al.~(in prep).

All measurements of the optical depth generated and used in this work are available in the paper and its online supplementary material, available on the journal's website as well as the first author's website\footnote{www.sarahbosman.co.uk/research}.

\section*{Acknowledgements}

This research has made use of NASA's Astrophysics Data System, and open-source projects including \texttt{ipython} \citep{ipython}, \texttt{scipy} \citep{scipy}, \texttt{numpy} \citep{numpy}, \texttt{astropy} \citep{astropy1,astropy2}, \texttt{scikit-learn} \citep{scikit-learn} and \texttt{matplotlib} \citep{matplotlib}.

SEIB, RAM, FW and MO acknowledge funding from the European Research Council (ERC) under the European Union’s Horizon $2020$ research and innovation programme (grant agreement No.~$740246$ ``Cosmic Gas''). AP acknowledges support from the ERC Advanced Grant INTERSTELLAR H2020/740120. AM acknowledges funding from the ERC under the European Union’s Horizon 2020 research and innovation programme (grant agreement No 638809 -- AIDA). The results presented here reflect the authors' views; the ERC is not responsible for their use.

LCK was supported by the European Union’s Horizon 2020 research and innovation programme under the Marie Sk\l odowska-Curie grant agreement No.~885990.

ACE acknowledges support by NASA through the NASA Hubble Fellowship grant $\#$HF2-51434 awarded by the Space Telescope Science Institute, which is operated by the Association of Universities for Research in Astronomy, Inc., for NASA, under contract NAS5-26555. F.W.~acknowledges support by NASA through the NASA Hubble Fellowship grant $\#$HST-HF2-51448.001-A awarded by the Space Telescope Science Institute, which is operated by the Association of Universities for Research in Astronomy, Incorporated, under NASA contract NAS5-26555.

GDB and YZ are supported by the National Science Foundation through grant AST-1751404. JFH
acknowledges support from the National Science Foundation under Grant No.~1816006.

RD, ERW and YQ acknowledge the Australian Research Council Centre of Excellence for All Sky Astrophysics in 3 Dimensions (ASTRO 3D), through project number CE170100013.

FB acknowledges support from the Australian Research Council through Discovery Projects (award DP190100252) and Chinese Academy of Sciences (CAS) through a China-Chile Joint Research Fund (CCJRF1809) administered by the CAS South America Center for Astronomy (CASSACA).

Based on observations collected at the European Southern Observatory under ESO programme 1103.A-0817(A).

Funding for the Sloan Digital Sky Survey IV has been provided by the Alfred P.~Sloan Foundation, the U.S.~Department of Energy Office of Science, and the Participating Institutions. SDSS acknowledges support and resources from the Center for High-Performance Computing at the University of Utah. The SDSS web site is \url{www.sdss.org}.

SDSS is managed by the Astrophysical Research Consortium for the Participating Institutions of the SDSS Collaboration including the Brazilian Participation Group, the Carnegie Institution for Science, Carnegie Mellon University, the Chilean Participation Group, the French Participation Group, Harvard-Smithsonian Center for Astrophysics, Instituto de Astrof\'isica de Canarias, The Johns Hopkins University, Kavli Institute for the Physics and Mathematics of the Universe (IPMU) / University of Tokyo, the Korean Participation Group, Lawrence Berkeley National Laboratory, Leibniz Institut f\"ur Astrophysik Potsdam (AIP), Max-Planck-Institut f\"ur Astronomie (MPIA Heidelberg), Max-Planck-Institut f\"ur Astrophysik (MPA Garching), Max-Planck-Institut f\"ur Extraterrestrische Physik (MPE), National Astronomical Observatories of China, New Mexico State University, New York University, University of Notre Dame, Observat\'orio Nacional / MCTI, The Ohio State University, Pennsylvania State University, Shanghai Astronomical Observatory, United Kingdom Participation Group, Universidad Nacional Aut\'onoma de M\'exico, University of Arizona, University of Colorado Boulder, University of Oxford, University of Portsmouth, University of Utah, University of Virginia, University of Washington, University of Wisconsin, Vanderbilt University, and Yale University.

The  Sherwood  simulation  was  performed  with  super-computer  time  awarded  by  the  Partnership  for  Advanced Computing in Europe (PRACE) $8$th call. This project also made use of the DiRAC High Performance Computing System (HPCS) and the COSMOS shared memory service at the University of Cambridge. These are operated on behalf of  the  Science  and  Technology  Facilities  Council  (STFC) DiRAC  HPC  facility.  This  equipment  is  funded  by  BIS National E-infrastructure capital grant ST/J005673/1 and STFC grants ST/H008586/1, ST/K00333X/1.

Calculations presented in this paper used resources of the National Energy Research Scientific Computing Center (NERSC), which is supported by the Office of Science of the U.S. Department of Energy under Contract No. DE-AC02-05CH11231.




\bibliographystyle{mnras}
\bibliography{bibliography} 




\appendix




\section{Alternative data binning}\label{app:length_binning}

\begin{figure*}
\includegraphics[width=\textwidth]{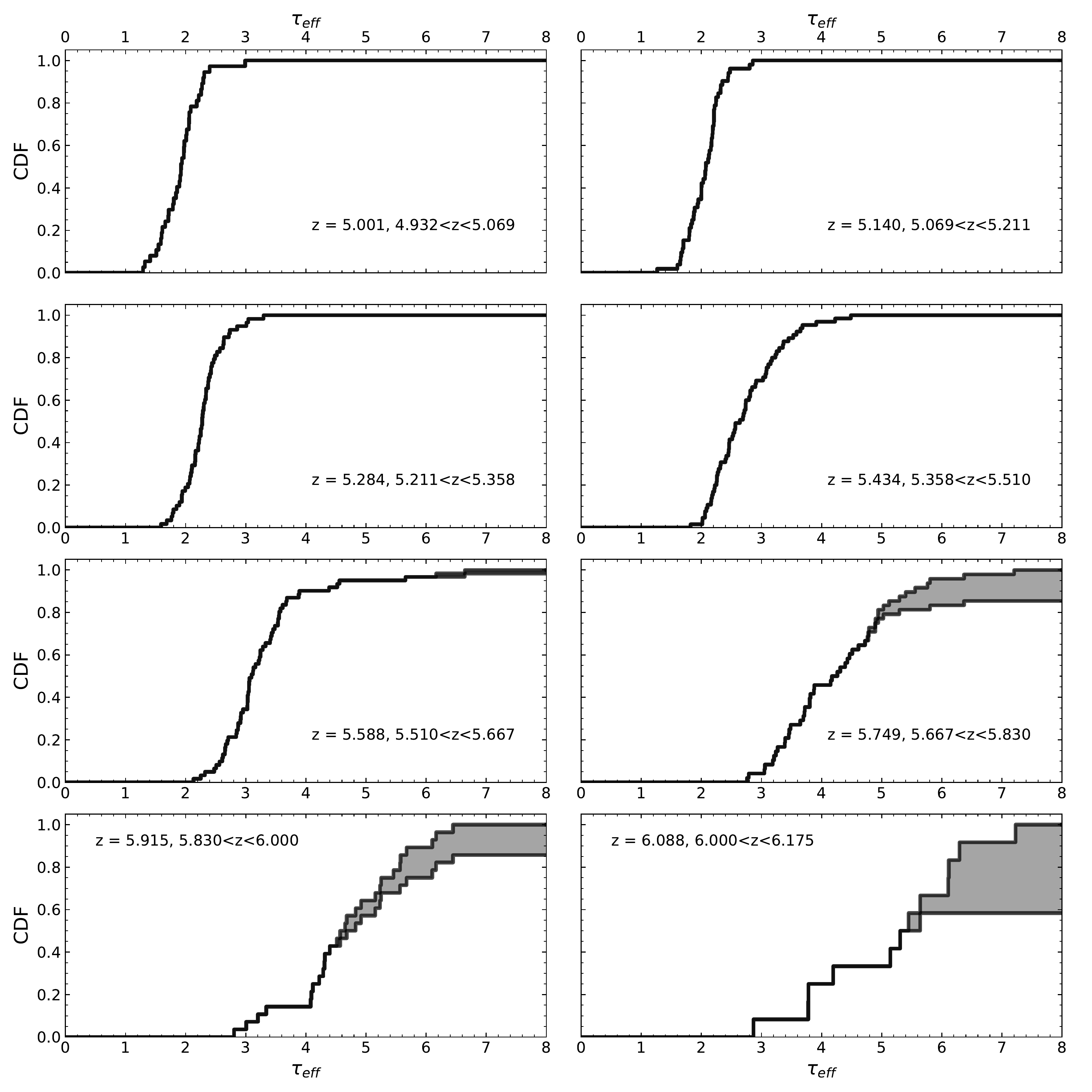}
\caption{Same as Figure~\ref{fig:dz}, but with fixed redshift intervals of constant length $\Delta L = 50$ cMpc/h as indicated in each panel. The qualitative evolution is unchanged: the first opaque troughs emerge at $z\sim 5.6$.}  \label{fig:app-50}
\end{figure*}

In this paper, we bin the data in fixed redshift intervals of equal size $\Delta z = 0.1$. 
Fixed redshift bins to equal comoving length may sometimes be more convenient for model comparison. In Figure~\ref{fig:app-50} we show a such a sub-division of the data between $4.9\lesssim z \lesssim 6.2$ in $8$ bins with fixed $\Delta L = 50$ cMpc/h. The mid-points and edges of each bin are given in each panel as well as in Table~\ref{table:app-50}. The average optical depths measured in this manner, as well as the qualitative evolution of the optical depth distribution, are fully consistent with those obtained in the paper's main body. The number of non-detections is slightly lessened due to averaging of the optical depth over a larger interval. Distributions of optical depths on different scales are expected to be sensitive to different optical effects. We make these distributions available as supplemental online material, as well as the distributions with intervals of $\Delta z = 0.05, 0.1, 0.2$ and $\Delta L = 30, 50, 70$ cMpc/h. Note that intervals $\Delta z \gtrsim 0.1$ are subject to variations of the mean flux $>1\sigma$ between their edges (Fig.~\ref{fig:means}).

\begin{table}
\centering
\begin{tabular}{l l l l c}
$z$ & $z_\text{min}$ & $z_\text{max}$ & $\left< F_{\text{Ly-}\alpha} \right> \ \ \ -1\sigma \ \ \ \ \ +1\sigma $ & $N_\text{los}$ \\
\hline
$5.000$ & $4.932$ & $5.069$ & $0.1545 \ -0.0085 \ +0.0080$ & $ 37 $ \\
$5.140$ & $5.069$ & $5.210$ & $0.1329 \ -0.0056 \ +0.0054$ & $ 52 $ \\
$5.284$ & $5.210$ & $5.357$ & $0.1073 \ -0.0043 \ +0.0046$ & $ 58 $ \\
$5.433$ & $5.357$ & $5.509$ & $0.0741 \ -0.0039 \ +0.0047$ & $ 65 $ \\
$5.588$ & $5.509$ & $5.667$ & $0.0458 \ -0.0032 \ +0.0031$ & $ 61 $ \\
$5.748$ & $5.667$ & $5.830$ & $0.0192 \ -0.0025 \ +0.0022$ & $ 48 $ \\
$5.915$ & $5.830$ & $5.999$ & $0.0126 \ -0.0031 \ +0.0028$ & $ 28 $ \\
$6.087$ & $5.999$ & $6.175$ & $0.0091 \ -0.0051 \ +0.0051$ & $ 12 $ \\
\end{tabular}
\caption{Mean \lal flux transmission at $4.9\lesssim z\lesssim 6.2$, measured in $\Delta L = 50$ cMpc/h bins with fixed redshift bounds $z_\text{min}$ and $z_\text{max}$. Uncertainties correspond to the $16$th and $84$th percentiles from bootstrap resampling. The measurement uncertainties on their own are a factor $5-10$ smaller than the bootstrap uncertainties quoted here. $N_\text{los}$ sightlines contribute to each measurement.}\label{table:app-50}
\end{table}

\section{$z=5.4$ distribution without most opaque sightlines}\label{app:wo}

\begin{figure*}
\includegraphics[width=\textwidth]{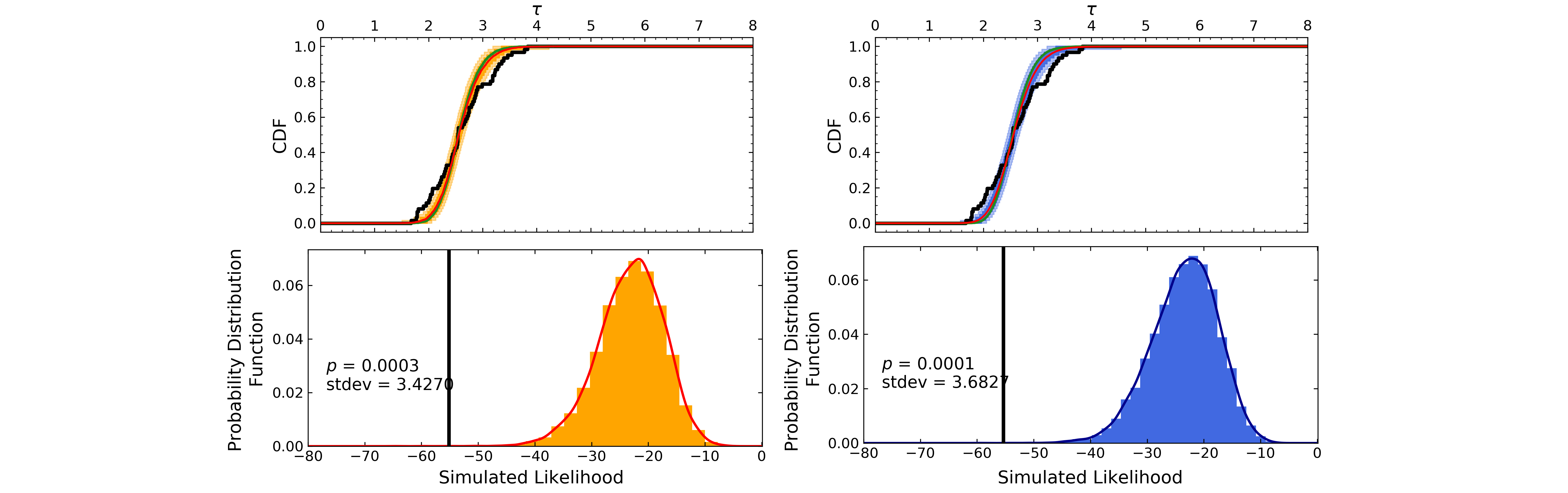}
\caption{Same as Figure~\ref{fig:Sherwood} but distributions at $z=5.4$ arbitrarily excluding the most opaque $3$ sightlines with $\tau>4$. Left: results for Sherwood. Right: results for Nyx. In both simulations, a tension at $>3\sigma$ remains even after removing the most opaque sightlines.}  \label{fig:wo}
\end{figure*}

To determine whether the tension at $z=5.4$ is due to outlier sightlines (potential DLAs), we arbitrarily remove the top $3$ most opaque sightlines and re-run the likelihood analysis. The results are shown in Figure~\ref{fig:wo}. While the tension is reduced compared to including the opaque sightlines, the tension remains above $3\sigma$ for both the Sherwood and Nyx simulations. The test therefore indicates that the entire shape of the \lal optical depth distribution at $z=5.4$, and not just a few outliers, drive the tension with homogeneous UVB models.



\bsp	
\label{lastpage}
\end{document}